\documentclass[structabstract]{aa}  
\usepackage{graphicx}
\usepackage{txfonts}
\usepackage{natbib}
\usepackage{hyperref}
\usepackage{siunitx}  
\usepackage{xcolor}
\newcommand{\HII}{\mbox{H\,{\sc ii}}}

\newcommand{\mci}[1]{\multicolumn{1}{c}{#1}}
\newcommand{\mcii}[1]{\multicolumn{2}{c}{#1}}
\newcommand{\mciii}[1]{\multicolumn{3}{c}{#1}}
\newcommand{\GG}{\mbox{$G_2$}}
\newcommand{\GGc}{\mbox{$G_2^\prime$}}
\newcommand{\GGG}{\mbox{$G_3$}}
\newcommand{\GGGc}{\mbox{$G_3^\prime$}}

\newcommand{\GGBP}{\mbox{$G_{\rm BP,2}$}}
\newcommand{\GGGBP}{\mbox{$G_{\rm BP,3}$}}
\newcommand{\GGRP}{\mbox{$G_{\rm RP,2}$}}
\newcommand{\GGGRP}{\mbox{$G_{\rm RP,3}$}}
\newcommand{\BPRP}{\mbox{$G_{\rm BP,3}-G_{\rm RP,3}$}}
\newcommand{\pic}{\mbox{$\varpi_{\rm c}$}}
\newcommand{\microas}{\mbox{$\mu$as}}
\newcommand{\sG}{\mbox{$\sigma_G$}}
\newcommand{\sX}{\mbox{$\sigma_X$}}
\newcommand{\Cstar}{\mbox{$C^*$}}
\newcommand{\pmra}{\mbox{$\mu_{\alpha *}$}}
\newcommand{\pmdec}{\mbox{$\mu_{\delta}$}}
\newcommand{\pmrag}{\mbox{$\mu_{\alpha *,{\rm g}}$}}
\newcommand{\pmdecg}{\mbox{$\mu_{\delta,{\rm g}}$}}
\newcommand{\pmlong}{\mbox{$\mu_{l*,{\rm g}}$ }}
\newcommand{\pmlatg}{\mbox{$\mu_{b,{\rm g}}$}}
\newcommand{\pmrac}{\mbox{$\mu_{\alpha *,{\rm c}}$}}
\newcommand{\pmdecc}{\mbox{$\mu_{\delta,{\rm c}}$}}
\newcommand{\rmu}{\mbox{$r_\mu$}}
\newcommand{\alphac}{\mbox{$\alpha_{\rm c}$}}
\newcommand{\deltac}{\mbox{$\delta_{\rm c}$}}
\newcommand{\Teff}{\mbox{$T_{\rm eff}$}}
\newcommand{\logd}{\mbox{$\log d$}}
\newcommand{\sigmas}{\mbox{$\sigma_{\rm s}$}}
\newcommand{\sigmai}{\mbox{$\sigma_{\rm int}$}}
\newcommand{\sigmae}{\mbox{$\sigma_{\rm ext}$}}
\newcommand{\pig}{\mbox{$\varpi_{\rm g}$}}
\newcommand{\spig}{\mbox{$\sigma_{\varpi_{\rm g}}$}}

\newcommand{\VO}[1]{Villafranca~O-{#1}}
\newcommand{\EBV}{\mbox{$E(4405-5495)$}}
\newcommand{\RV}{\mbox{$R_{5495}$}}
\newcommand{\AV}{\mbox{$A_{5495}$}}
\newcommand{\AG}{\mbox{$A_G$}}
\newcommand{\MG}{\mbox{$M_G$}}
\newcommand{\Msun}{\mbox{M$_\odot$}}
\newcommand{\chired}{\mbox{$\chi^2_{\rm red}$}}
\newcommand{\rc}{\mbox{$r_{\rm c}$}}
\newcommand{\fc}{\mbox{$f_{\rm c}$}}
\newcommand{\fb}{\mbox{$f_{\rm b}$}}

\begin{document}

   \title{An in-depth analysis of the differentially expanding \linebreak 
          star cluster Stock 18 (\VO{036}) \linebreak
          using \textit{Gaia} DR3 and ground-based data}
   \titlerunning{An analysis of Stock 18 using \textit{Gaia} DR3 and ground-based data}
   \authorrunning{Ma\'{\i}z Apell\'aniz et al.}


   \author{J. Ma\'{\i}z Apell\'aniz\inst{1}
          \and
          A. R. Youssef\inst{2}
          \and
          M. S. El-Nawawy\inst{2}\thanks{Deceased.}
          \and
          W. H. Elsanhoury\inst{3,4}
          \and
          A. Sota\inst{5}
          \and
          \linebreak
          M. Pantaleoni Gonz\'alez\inst{1,6}
          \and
          A. Ahmed\inst{2}
          }

   \institute{Centro de Astrobiolog\'{\i}a, CSIC-INTA. Campus ESAC. 
              C. bajo del castillo s/n. 
              E-\num[detect-all]{28692} Villanueva de la Ca\~nada, Madrid, Spain.\\
              \email{jmaiz@cab.inta-csic.es}
         \and
             Astronomy, Space Science, and Meteorology Department.
             Faculty of Science, Cairo University. 
             12613, Giza, Egypt.
         \and
             Department of Physics.
             College of Science, Northern Border University.
             Arar, Saudi Arabia.
         \and
             Department of Astronomy.
             National Research Institute of Astronomy and Geophysics (NRIAG). 
             11421, Helwan, Cairo, Egypt. 
         \and
             Instituto de Astrof\'{\i}sica de Andaluc\'{\i}a (IAA), CSIC. 
             Glorieta de la Astronom\'{\i}a s/n. E-\num[detect-all]{18008} Granada, Spain
         \and
             Departamento de Astrof{\'\i}sica y F{\'\i}sica de la Atm\'osfera. 
             Universidad Complutense de Madrid. 
             E-\num[detect-all]{28040} Madrid, Spain. 
             }

   \date{Received 19 April 2024; accepted 22 May 2024}

 
  \abstract
   {The Villafranca project is combining \textit{Gaia} data with ground-based surveys to analyze Galactic stellar groups (clusters, associations, or parts thereof) with OB stars.}
   {We want to analyze the poorly studied cluster Stock~18 within the Villafranca project, as it is a very young stellar cluster with a symmetrical and compact \HII\ region around it, Sh~2-170, so it is likely to provide insights into the structure and dynamics of such objects at an early stage of their evolution.}
   {We use \textit{Gaia} astrometry, photometry, spectrophotometry, and variability data and ground-based spectroscopy and imaging to determine the characteristics of Stock~18. We use them to analyze its core, massive-star population, extinction, distance, membership, internal dynamics, density profile, age, IMF, total mass, stellar variability, and Galactic location.}
   {Stock~18 is a very young ($\sim 1.0$~Ma) cluster located at a distance of $2.91\pm0.10$~kpc dominated by the GLS~\num{13370} system, whose primary (Aa) is an O9~V star. We propose that Stock~18 was in a very compact state ($\sim$0.1~pc) about 1.0~Ma ago and that most massive stars were ejected at that time without significantly affecting the less massive stars as a result of multi-body dynamical interactions. Different age estimates also point out towards an age close to 1.0~Ma, indicating that the dynamical interactions took place very soon after massive star formation. Well defined expanding stellar clusters have been observed before but none as young as this one. If we include all of the stars, the initial mass function is top heavy but if we discard the ejected ones it becomes nearly canonical. Therefore, this is another example in addition to the one we previously found (the Bermuda cluster) of (a) a very young cluster with an already evolved present day mass function (b) that has significantly contributed to the future population of free-floating compact objects. If confirmed in more clusters, the number of such compact objects may be higher in the Milky Way than previously thought. Stock~18 has a variable extinction with an average value of \RV\ higher than the canonical one of 3.1. We have discovered a new visual component (Ab) in the GLS~\num{13370} system. The cluster is above our Galactic mid-plane, likely as a result of the Galactic warp, and has a distinct motion with respect to its surrounding old population, which is possibly an influence of the Perseus spiral arm.}
   {}


   \maketitle
%

\section{Introduction}

$\,\!$\indent The analysis of open star clusters provides a window into our comprehension of stellar evolution and the structure and evolution of the Milky Way (MW) thin disc, as they are foundational units of galaxies \citep{gilmore2012gaia}.  The youngest open clusters, those with ages of only several Ma, are particularly interesting as their conditions are still similar to the initial ones, shedding light into the process of star formation, and as they possess present day mass functions (PDMFs) only slightly altered from the initial one (or IMF). 

\begin{table*}
\centering
\caption{Results from previous studies of Stock~18.}
\label{Stock18_results}
\setlength\tabcolsep{3pt}
\begin{tabular}{ccccccccccc}
\hline
$\alpha_{\rm J2000}$ & $\delta_{\rm J2000}$ & $r$       & \rc       & Age   & $d$   & $E(B-V)$ & $\mu_\alpha\cos\delta$ & $\mu_\delta$       & $N$     & Ref.  \\
($^{\circ}$)         & ($^{\circ}$)         & (\arcmin) & (\arcmin) & (Ma)  & (kpc) & (mag)    & (mas $\rm a^{-1}$)     & (mas $\rm a^{-1}$) & (stars) &       \\
\hline
0.395 & 64.624 & 8.5 & ---       & ---       & 2.60$\pm$0.27       & 0.56-0.70 & ---                & ---                        &   5 & 1  \\
0.400 & 64.623 & 6.0 & 0.37      & 130       & 1.25$\pm$0.08       & 0.69      & ---                & ---                        & 259 & 2  \\
0.450 & 64.614 & --- & ---       & ---       & 10.3                & ---       & ---                & ---                        &  44 & 3  \\
0.398 & 64.625 & 4.8 & 0.71      & 480       & 0.77                & 0.177     & $-$3.59            & $-$1.15                    & --- & 4  \\ 
0.405 & 64.625 & 4.0 & ---       & ---       & ---                 & ---       & $-$2.78$\pm$1.59   & $-$0.15$\pm$0.84           & 109 & 5  \\
0.397 & 64.625 & --- & ---       & 480       & 0.77                & 0.177     & ---                & ---                        & --- & 6  \\
0.405 & 64.625 & 4.0 & ---       & 130       & 1.24                & 0.71      & $-$2.47$\pm$0.58   & \phantom{$-$}0.42$\pm$0.33 &  72 & 7  \\
0.399 & 64.625 & --- & ---       & ---       & $3.1^{+1.4}_{-0.7}$ & ---       & $-$2.692$\pm$0.020 & $-$0.587$\pm$0.020         &  26 & 8  \\ 
0.399 & 64.625 & --- & ---       &  13       & 2.86                & ---       & $-$2.692$\pm$0.075 & $-$0.587$\pm$0.076         &  22 & 9  \\
\hline
0.393 & 64.626 & 9.0 & ---       & 0.25-0.50 & 2.2$\pm$0.4         & 0.70-0.90 & ---                & ---                        & --- & 10 \\ 
0.396 & 64.627 & 3.5 & 0.30-0.56 & 6$\pm$2   & 2.8$\pm$0.2         & 0.70-0.90 & ---                & ---                        & --- & 11 \\ 
0.400 & 64.625 & 2.7 & 0.83      & 0-7       & 3.1$\pm$0.2         & 0.7       & $-2.62$            & $-0.49$                    &  86 & 12 \\
\hline \\
\end{tabular}

References: (1) \cite{russeil2007revised}; (2) \cite{Bukoetal11}; (3) \cite{BuckFroe13}; (4) \cite{kharchenko2013global}; \\
(5) \cite{dias2014proper}; (6) \cite{joshi2016study}; (7) \cite{sampedro2017multimembership}; (8) \cite{cantat2020clusters}; \\
(9) \cite{cantat2020painting}; (10) \cite{roger2004sharpless}; (11) \cite{bhatt2012stellar}; (12) \cite{sinha2020variable}. 
\end{table*}

The European Space Agency mission \textit{Gaia} has revolutionized the study of open star clusters by providing astrometric, photometric, and spectroscopic information for almost $2\cdot 10^9$ sources. We are combining \textit{Gaia} information with ground-based data to conduct the Villafranca project, a study of Galactic OB groups in the solar neighborhood. An OB group is a stellar ensemble born from a single cloud that may be bound (a cluster) or unbound (an association or part thereof) and that is massive and young enough to still contain one or (more typically) several OB stars. For the latter, we use the classical definition of a star with a spectral type earlier than B2 for dwarfs (luminosity class V), B5 for giants (luminosity class III), and B9 for supergiants (luminosity class I). The Villafranca project uses \textit{Gaia} astrometry, photometry, and spectroscopy to derive the properties of stellar OB groups with the help of ground-based spectroscopic and photometric surveys. The first ones, such as the Galactic O-Star Spectroscopic Survey (GOSSS, \citealt{Maizetal11}) and LiLiMaRlin \citep{Maizetal19a} are used to characterize the OB stars in the groups and are being combined into an umbrella project called the Alma Luminous Star (ALS) survey \citep{Reed03,Pantetal21} that is analyzing tens of thousands of Local Group massive stars. The second ones, such as 2MASS \citep{skrutskie2006two} and GALANTE \citep{Maizetal21d}, are used to determine properties such as extinction for the stars in the cluster or association. In the prototype Villafranca paper \citep{Maiz19} we presented the method used to determine the group characteristics and analyzed the first two groups with O stars. In the two major Villafranca papers published so far (\citealt{Maizetal20b} or paper~I and \citealt{Maizetal22a} or paper~II) we analyzed a total of 26 stellar groups with O stars (\VO{001} to \VO{026}) and their results have been used in subsequent papers to propose the phenomenon of orphan clusters \citep{Maizetal22b} and to recalibrate \textit{Gaia}~EDR3 (early third data release) astrometry \citep{Maizetal21c,Maiz22}. Since then, we have analyzed three additional Villafranca groups with O stars in \citet{Neguetal22,Putketal23,Ansietal23} and we are currently writing the next major paper of the series (Molina-Lera et al. in prep. or paper III), centered on the groups in Carina~OB1. Counting all, we have studied 35 Galactic groups with O stars and another 5 with massive B stars.

\begin{figure*}
 \centerline{\includegraphics*[width=0.49\linewidth]{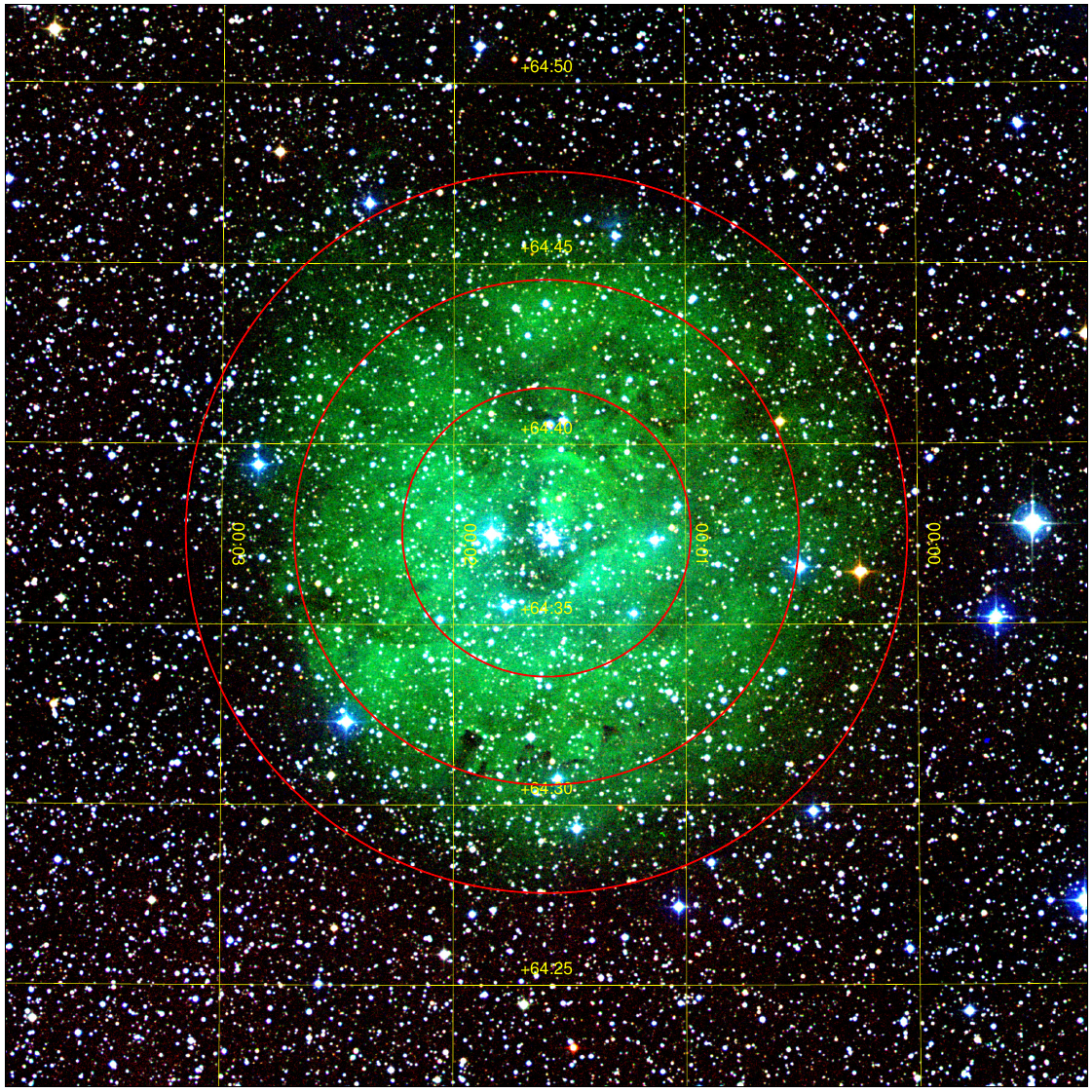} \
             \includegraphics*[width=0.49\linewidth]{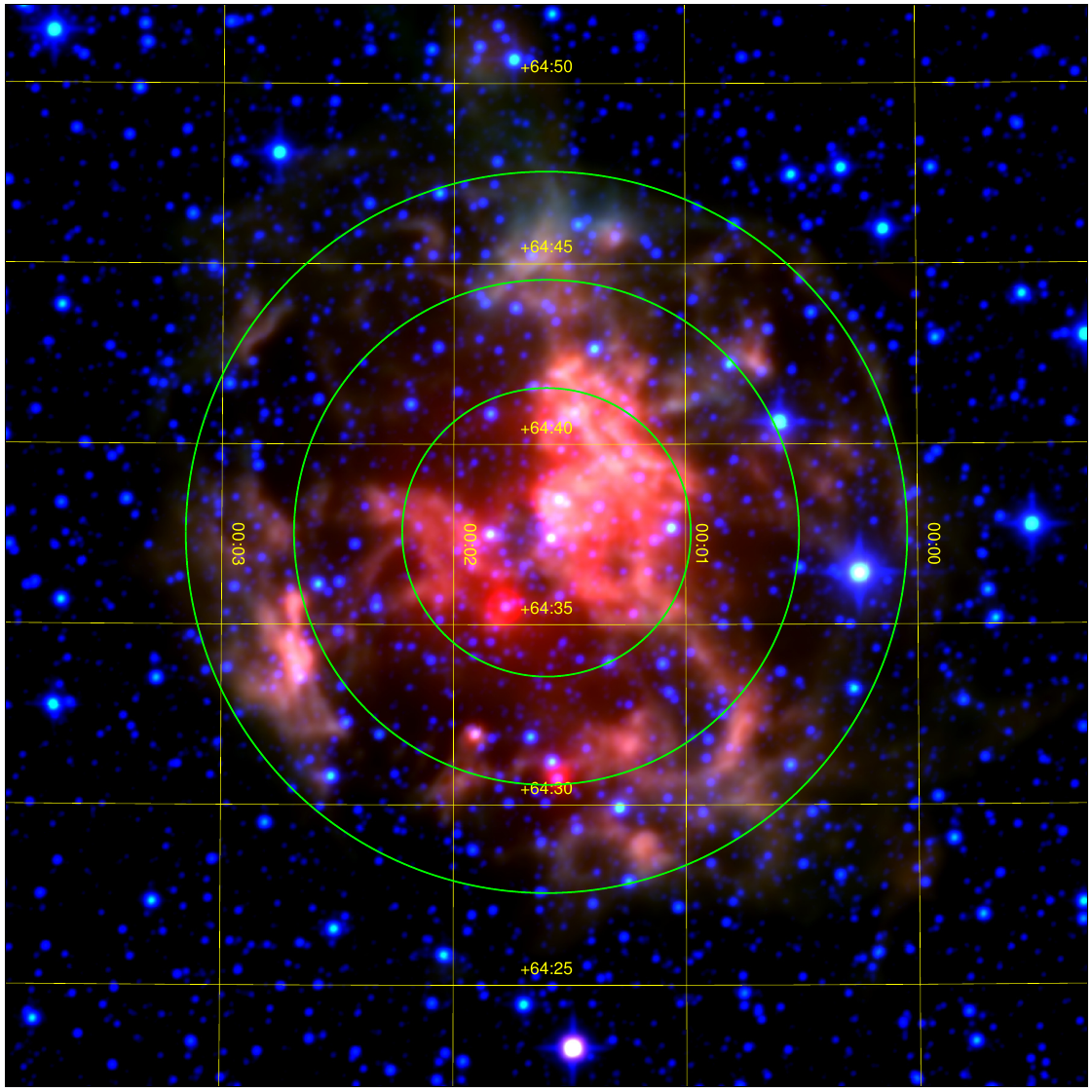}}
 \caption{(left) Three-channel DSS2 RGB image of Stock~18 with the red channel 
           corresponding to the NIR survey, the green channel to the Red survey, 
           and the blue channel to the Blue survey. (right) WISE RGB image of 
           Stock~18 with the red channel corresponding to the W4 band, the green 
           channel to a sum of the W3 and W2 bands, and the blue channel to the 
           W1 band. In both cases the intensity scale is logarithmic and circles
           with radii of 4\arcmin, 7\arcmin, and 10\arcmin\ centered on the
           cluster are plotted.}
 \label{DSS2_WISE}   
\end{figure*}

Stock~18 is one of the 21 open clusters observed by J\"urgen Stock \citep{macconnell12006homage} in the early 1950s. It is included in a number of large catalogs that have derived information about open clusters \citep{russeil2007revised,Bukoetal11,BuckFroe13,kharchenko2013global,dias2014proper,joshi2016study,sampedro2017multimembership,cantat2020clusters,cantat2020painting} but has only been studied in depth by \citet{roger2004sharpless}, \citet{bhatt2012stellar} and \citet{sinha2020variable}. A summary of the results from those papers is given in Table~\ref{Stock18_results}. 

A look at Table~\ref{Stock18_results} reveals a strong disparity in ages and distances in the first group (large catalogs) and a much better agreement in the second (dedicated studies). The disparity has been addressed in previous studies (\citealt{netopil2015comparative} and paper I) and is the result of several factors. The most important one is that large surveys have to make simplifying assumptions, many times ignoring significant clues about age and other parameters. Paper I also argued in favor of the positive role of \textit{Gaia} but we note here that two of the dedicated papers were published before \textit{Gaia} data became available and were still able to provide a good distance (within 2 sigmas of the value we derive below). Along the same lines, \citet{mayer1973photometry} determined a distance of 2.99~kpc to GLS~\num{13370}~A (=LS~I~$+64$~11~A), the only O star in the cluster (as we will see below), which is also a good value. The real problem with the distance appears to be its relationship with the age measurement, as several of the large surveys severely overestimate the latter. 

The (sometimes ignored) youth of Stock~18 is readily apparent by its association with the \HII\ region Sh~2-170 \citep{Shar59}, see Fig.~\ref{DSS2_WISE}. Sh~2-170 is a relatively compact, symmetrical, and filled \HII\ region with a single O star located very close to its geometrical center. The WISE image, with W4 tracing the warm dust, reveals just a small cavity around the O star, a sign of extreme youth. A detailed analysis of the \HII\ region by \citet{roger2004sharpless} led to an age estimate of just 0.25-0.50~Ma, counting since the activation of the O star. The analysis of the PMS (pre-main sequence) population by \citet{sinha2020variable} shows a peak at an age of 1~Ma with some stars as old as 7~Ma. The results in the two papers are consistent, as PMS models cannot easily discriminate age differences smaller than 1~Ma and star-formation on scales of $\sim$10~pc does not occur instantaneously: it is quite typical to have a cluster surrounded by a halo (or association) with a spread of ages of several Ma. 

The next section describes our data and methods and the next ones are each dedicated to a different aspect of our analysis of Stock~18: the cluster core; the massive-star population; extinction; cluster membership and distance; internal cluster dynamics and structure; age and variability; IMF and cluster mass; and Galactic location. We conclude with a summary of results, some future lines of research, and an appendix with a glossary of terms.

\section{Data and methods}
\label{data-sect}

\subsection{ALS spectroscopy}

$\,\!$\indent Most analyses of stellar clusters with \textit{Gaia} are based on data provided by the mission alone, perhaps with the additional help of other photometric surveys such as 2MASS. The problem of doing so with young groups with OB stars is that they are less ``democratic'' than older clusters, as the (typically few) massive stars play a disproportionately large role in the ionizing flux, stellar winds, and stellar dynamics compared to the far more numerous intermediate- and low-mass stars. In a sense, OB groups are defined by the OB stars themselves and the rest of the stars just tag along for the trip. The OB phase of a stellar group is short lived but it is the most important, as the events that take place there determine whether the rest of the stars remain bound in a cluster for the rest (or at least a significant fraction) of their lives. To study OB stars in detail, spectroscopy is needed due to the relatively poor information content of their photometry. For those reasons, in the Villafranca project we start the analysis of each group using spectroscopic information from its most outstanding denizens.

In the Villafranca project we have so far used two types of spectroscopy: Intermediate-resolution ($R\sim 2500$) from GOSSS and high-resolution ($R\ge \num{25000}$) from LiLiMaRlin. In this paper we use four GOSSS long-slits obtained with the OSIRIS spectrograph at the 10.4~m Gran Telescopio de Canarias (GTC). In each case, the slit was aligned to include two stars in Stock~18 for an expected total number of eight. As it turned out, the slit used to observe GLS~\num{13370}~A also included the nearby B companion (separation of 2\farcs1 and $\Delta \GGG$ of 2.5~mag) and we were able to use our well tested spatial deconvolution techniques \citep{Maizetal18a,Maizetal21b} to separate the visual secondary. Therefore, the total number of spectra was increased to nine (Tables~\ref{spclas}~and~\ref{Gaia_stars}). In a new paper of the ALS series (Pantaleoni Gonz\'alez et al. in prep.) we are merging the results of GOSSS and LiLIMaRlin into a single project that will be the successor of GOSC (Galactic O-Star Catalog, \citealt{Maizetal04b}), so from now on we will just use the term ALS spectroscopy to refer to the GTC spectroscopic data in this paper. We note that in the new ALS paper we change the nomenclature of the confirmed Galactic members of the catalog from the existing e.g. ALS~\num{13370} (with ALS standing for Alma Luminous Star) to GLS~\num{13370} (with GLS standing for Galactic Luminous Star), as some objects are being dropped from the catalog for not being massive stars and we are also including LMC and SMC massive stars (as LLS and SLS, respectively).

We derived spectral classifications using MGB \citep{Maizetal12,Maizetal15b}, a code that compares the observed spectra with a high-quality 2-D grid (spectral subtype and luminosity class). The code allows the user to fit the rotation index\footnote{Slowly-rotating stars have no index and increasing $v\,\sin i$ (or broadening the lines by other mechanism such as unresolved binarity) leads to subsequent indices of (n), n, nn, and nnn.} and to combine different grid spectra to derive spectral classification for SB2 systems. In this paper we use a new spectral classification grid (Ma\'{\i}z Apell\'aniz et al. in prep.) that significantly improves in coverage and grid sampling upon the previous grid of \citet{Maizetal16}.

\subsection{\textit{Gaia}~(E)DR3 coordinates, parallaxes, proper motions, magnitudes, and spectrophotometry}

$\,\!$\indent As in paper II, we downloaded from VizieR the \textit{Gaia}~EDR3 \citep{brown2021gaia} coordinates, parallaxes, proper motions, and magnitudes of the sources in the region of Stock~18. Note that the \textit{Gaia} epoch-averaged photometry and astrometry does not change between EDR3 and DR3 \citep{vallenari2023gaia}. In addition, we also downloaded the magnitudes from \textit{Gaia}~DR2 \citep{Rieletal18} and the \textit{Gaia}~DR3 XP spectrophotometry for selected sources. Each type of data was recalibrated, as described next.

We derived corrected parallaxes from the catalogued ones by applying the zero point (ZP) values:

\begin{equation}
\pic = \varpi - {\rm ZP},
\label{pic}
\end{equation}

\noindent as described in \citet{Maiz22}, with the values for ZP derived there using the results from paper II. The external parallax uncertainties were derived from the internal or catalog uncertainties using Eqn.~1 from \citet{Fabretal21a}:

\begin{equation}
\sigmae = \sqrt{k^2 \sigma_{\rm int}^2 + \sigma_{\rm s}^2}, 
\label{sigmae}  
\end{equation}

\noindent with the $k$ multiplicative values from \citet{Maiz22} and the systematic uncertainty \sigmas\ of 10.3~\microas\ from \citet{Maizetal21c}. We point out the need to use external uncertainties instead of internal (catalog) ones because $k$ is significantly greater than 1.0 in most cases of interest \citep{Fabretal21a}: \textit{Papers that use catalog Gaia DR3 parallaxes directly significantly underestimate parallax or distance uncertainties.} Also, it is important to note that the angular covariance should be taken into account when combining cluster member parallaxes to derive cluster distances (\citealt{Maizetal21c}, see below).

The proper motions for bright stars were corrected using \citet{CanGBran21}. Equation~\ref{sigmae} was also applied to the proper motion uncertainties, in this case using the $k$ values from \citet{Maiz22} and the \sigmas\ of 23~\microas/a from \citet{lindegren2021gaia}. We note, however, that when calculating relative (not absolute) proper motions in a small region of the sky one should set \sigmas\ to zero, as proper motions separated by a few arcminutes are highly correlated. 

\begin{figure*}
 \centerline{\includegraphics*[width=0.467\linewidth]{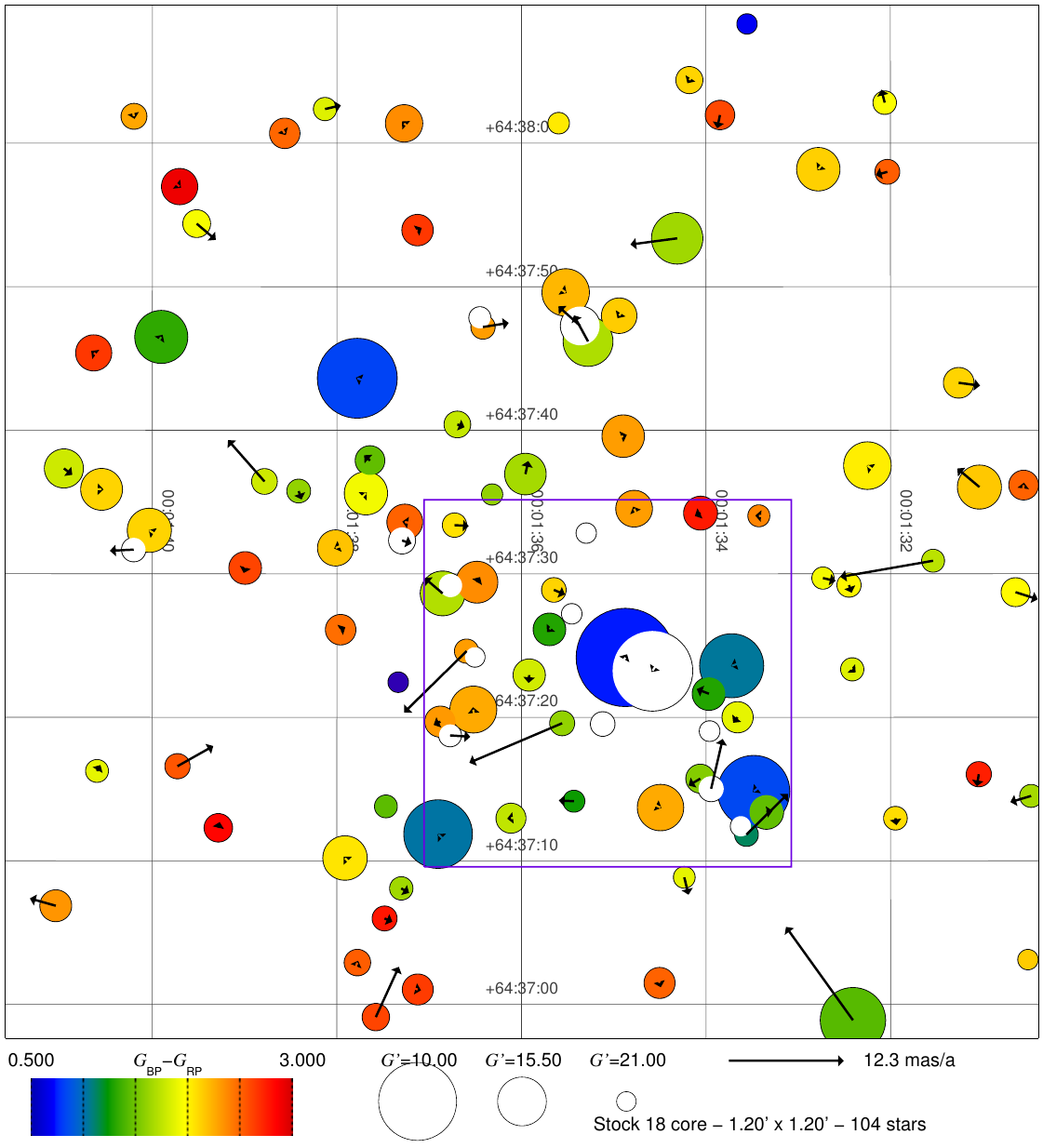} \
             \includegraphics*[width=0.513\linewidth]{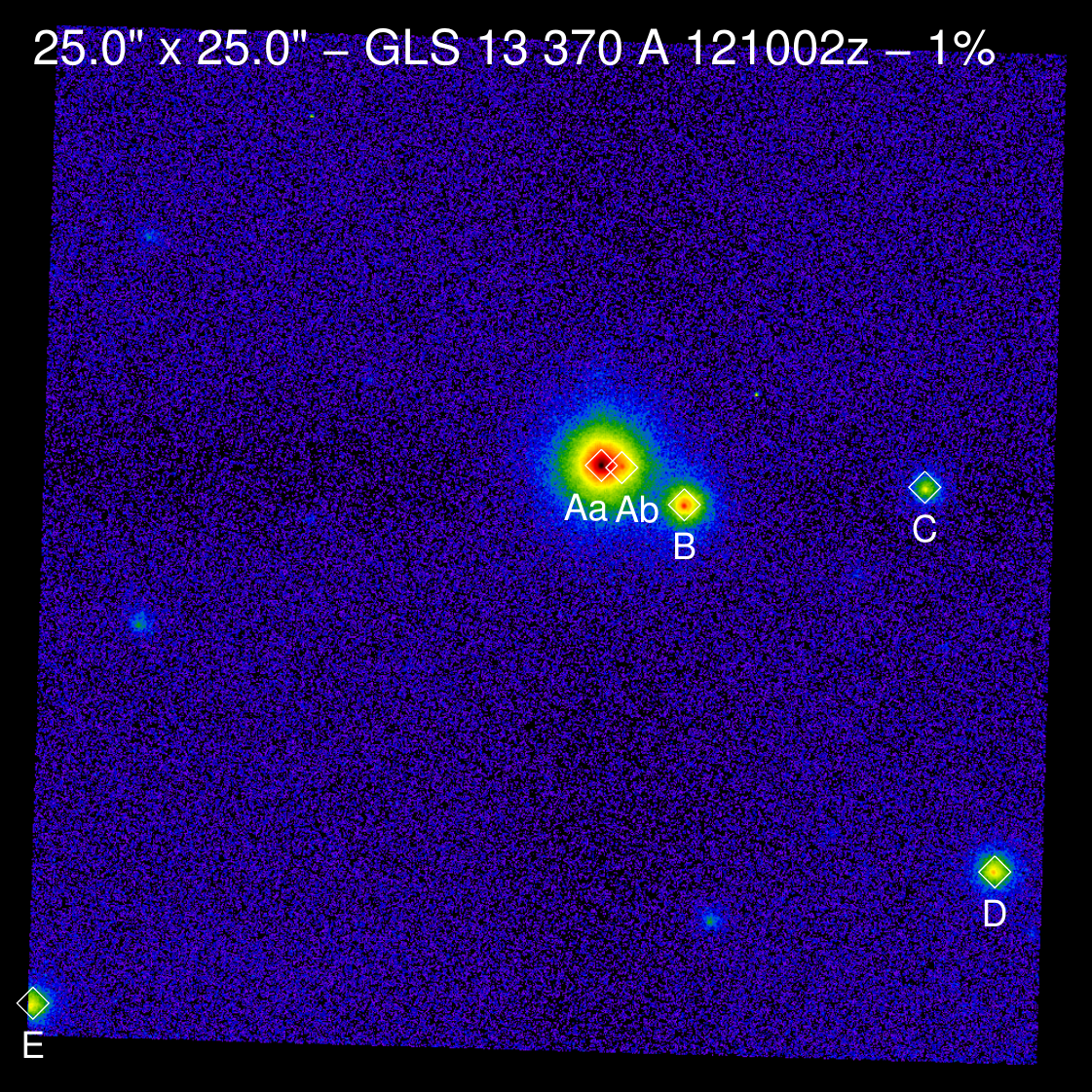}}
 \caption{(left) \textit{Gaia} chart for all stars (members and non-members alike) in the central 
          $1.2\arcmin\times1.2\arcmin$ of Stock~18. Symbol size represents \GGGc\ magnitude, symbol color 
          \BPRP, and arrows proper motion. Stars without \BPRP\ are shown without a color. 
          The proper motions have the mean value 
          derived for sample~1 subtracted. The purple square indicates the region shown in the right 
          panel. (right) AstraLux $z$ lucky image of the GLS~\num{13370} system, indicating the five 
          WDS components with A divided into Aa and Ab.  The intensity scale is 
          logarithmic to show both bright and faint sources.}
 \label{Stock_18_core}   
\end{figure*}

In \citet{MaizWeil18} we recalibrated the \textit{Gaia} DR2 \GG+\GGBP+\GGRP\ photometry using HST spectrophotometry. In a subsequent paper (Weiler et al. in prep.) we have done the same for the \textit{Gaia} EDR3 \GGG+\GGGBP+\GGGRP\ and we have used new data to improve upon the previous DR2 calibration. In the second paper we show that small corrections have to be applied to both \GG\ and \GGG\ (yielding \GGc\ and \GGGc) and that the DR2 and EDR3 photometric systems are different enough for information to be contained in colors derived from different DRs (e.g. $\GGGBP-\GGBP$). Below we describe how we have used the combined photometry to derive information about the extinction towards Stock~18.

\subsection{Cluster membership and distance}

$\,\!$\indent In this paper we add Stock~18 to our list of Galactic stellar groups with O stars as its 36th member, that is, \VO{036}.  The Villafranca project selects cluster members using a technique that is somewhat different from other projects that use \textit{Gaia} for the same purpose. The reader is referred to previous papers for details and to the appendix for definitions, here we just provide a brief description.

Rather than using an automatic procedure, we first identify the OB stars in the cluster (preferably from spectroscopy, see above) and we use them to define the characteristic position, proper motions, and extinction of the stellar group. We use those to select an initial guess for the cluster center and radius (\alphac, \deltac, $r$), central proper motion and radius (\pmrac, \pmdecc, $r_\mu$), color range with respect to a reference isochrone [$\Delta(\BPRP)$], and possibly other quantities such as maximum RUWE, \Cstar, and parallax uncertainty \sigmae. A code written in IDL is then used to calculate the distance $d$ iteratively excluding those objects whose external parallaxes are more than 3 \sigmae\ away from the group parallax \pig, a step in which it is crucial to use the correct uncertainties (see above). The code allows to interactively see which stars are included and which ones are excluded, change the parameters accordingly, and iterate the procedure until the desired result is obtained. Such a procedure has the advantage of being flexible, allowing to test the existence of structures such as double cores and halos, ensure that stars that are known to be cluster members are included, and to vary the degree of purity/completeness of the final result, possibly presenting different options. This is especially relevant for young clusters, whose complex morphology may be missed by automatic procedures. It has the disadvantage of being time consuming, limiting its use to relatively small cluster samples.

As we have already mentioned, one needs to take into account that \textit{Gaia} parallaxes have a small but significant angular covariance, with an oscillatory component with a wavelength of $\sim 1^{\circ}$ and a large-angle component \citep{Lindetal18a,lindegren2021gaia,Maizetal21c}. Therefore, when combining parallaxes of a set of objects assumed to be at the same distance, one has to account for that effect summing over all pairs of stars (see \citealt{Lindetal18b} for a recipe on how to do that). In practical terms, for a cluster with a significant number of stars and an angular size of several arcminutes, the angular covariance dominates the group parallax uncertainty determined from \textit{Gaia}~EDR3 to a value that is the distance in kpc expressed as a percentage. In other words, a cluster at a distance of $\sim$1~kpc will have a parallax (or distance) uncertainty of $\sim$1\%, one at $\sim$3~kpc an uncertainty of $\sim$3\%, and so on.

As we are dealing with a young stellar cluster, we cannot use a general Galactic distance prior, which is dominated by old stars with a very different spatial distribution. Instead, we use the prior for OB stars derived by \citet{Maiz01a,Maiz05c} updated with the parameters from \citet{Maizetal08a} and selecting the disk component.

\begin{table*}
\caption{GLS~\num{13370} component pairs measurements from this paper (AstraLux) and the literature. The WDS measurements are the last ones in that database.}
\label{GLS_13_370_pairs}
\centerline{
\begin{tabular}{lrrrrrrrrr}
\hline
Name  & \multicolumn{3}{c}{AstraLux}                & \multicolumn{3}{c}{\textit{Gaia}}            & \multicolumn{3}{c}{WDS}                     \\
      & \mci{sep.} & \mci{PA} & \mci{$\Delta z$}    & \mci{sep.} & \mci{PA} & \mci{$\Delta \GGGc$} & \mci{sep.} &  \mci{PA} & \mci{$\Delta m$}   \\
      & \mci{(\arcsec)} & \mci{(deg)} & \mci{(mag)} & \mci{(\arcsec)} & \mci{(deg)} & \mci{(mag)}  & \mci{(\arcsec)} & \mci{(deg)} & \mci{(mag)} \\
\hline
Aa,Ab &       0.49 &    267.4 &                3.91 & \mci{---} & \mci{---} &            \mci{---} &  \mci{---} & \mci{---} &          \mci{---} \\
Aa,B  &       2.10 &    243.9 &                2.52 &      2.11 &     243.7 &                 2.50 &        2.1 &       244 &                2.5 \\
Aa,C  &       7.43 &    265.7 &                4.63 &      7.43 &     265.5 &                 4.82 &        7.4 &       266 &                5.2 \\
Aa,D  &      12.91 &    223.9 &                3.69 &     12.93 &     223.8 &                 3.70 &       12.9 &       224 &                3.6 \\
Aa,E  &      17.94 &    133.4 &                4.14 &     17.94 &     133.3 &                 4.18 &       17.9 &       133 &                4.1 \\
\hline
\end{tabular}
}
\end{table*}
\subsection{Lucky imaging}

$\,\!$\indent We used the AstraLux instrument at the 2.2~m Calar Alto telescope to obtain lucky imaging of the core of Stock~18 using the $z$ band on 2 October 2012. The observing program and general characteristics of the data are given in \citet{Maiz10a} and the PSF-fitting procedure of the images and examples are given in \citet{SimDetal15a,Maizetal19b}. The right panel of Fig.~\ref{Stock_18_core} shows the observed field.

\subsection{Additional photometry and CHORIZOS analysis}

$\,\!$\indent For the stars with ALS spectroscopy we have collected their 2MASS photometry \citep{skrutskie2006two}. GLS~\num{13370} is unresolved in 2MASS and only partially resolved in \textit{Gaia}~DR3, as A has photometry in the three bands but B only in \GGG. We have tested that the three \textit{Gaia}~DR3 magnitudes are incompatible among them (in the sense of belonging to a typical extinguished O-type SED) and that it is likely that the \GGGBP\ and \GGGRP\ are intermediate values between those of an isolated A and a combined A,B system. Furthermore, as we will see below, there is at least a third hidden component in the system. Such a problematic \textit{Gaia}~DR3 photometry is not strange for a partially resolved multiple system such as GLS~\num{13370}. For those reasons, in the procedure described in the next paragraph, we treat GLS~\num{13370} as unresolved with a \GGGc\ magnitude intermediate between those of A and A,B and in the CHORIZOS analysis below we substitute the \textit{Gaia} \GG+\GGBP+\GGGBP+\GGRP+\GGGRP\ bands by Tycho-2 $B_{\rm T}+V_{\rm T}$ \citep{Hogetal00a}, see \citet{MaizBarb18} for similar examples of this procedure.

We use the SED-fitting code CHORIZOS \citep{Maiz04c} to calculate the extinction properties of the stars with ALS spectroscopy. We use the SED grid of \citet{Maiz13a}, the family of extinction laws of \citet{Maizetal14a}, and the photometric calibration of \citet{Maiz05b,Maiz06a,Maiz07a,MaizWeil18,MaizPant18} plus the aforementioned recalibration of \textit{Gaia}~DR3 photometry (Weiler et al. in prep.). As shown in \citet{MaizBarb18} and \citet{Maizetal21a}, the family of extinction laws of \citet{Maizetal14a} provides a better fit to optical-NIR photometry of Galactic OB stars than other existing alternatives. See \citet{Maiz24} for an explanation of the differences between monochromatic extinction quantities such as \EBV\ and \RV\ (see appendix and \citealt{Maiz24}) and their band-integrated equivalents such as $E(B-V)$ and $R_V$. For each star we fix \Teff\ from the spectral type using the \citet{Holgetal18} scale and \logd\ from the distance to Stock~18 determined below and we leave as free parameters the luminosity class \citep{Maiz13a}, a photometric equivalent to the spectroscopy luminosity class ranging from 0.0 (hypergiants) to 5.5 (ZAMS), and two extinction parameters for amount [\EBV] and type [\RV] of dust. As photometric information we use the nine-filter system $\GGBP+\GG+\GGRP+\GGGBP+\GGG+\GGGRP+J+H+K$, except for GLS~\num{13370}~A,B where, as already mentioned, we use Tycho-2 photometry and \GGG\ in the optical for a total of six filters including the three 2MASS ones.

\subsection{Variability information}

$\,\!$\indent We have cross-matched the sample in this paper with \citet{Maizetal23} to obtain the three-band \textit{Gaia}~DR3 photometric dispersions of the stars with $\GGG < 17$~mag and five-parameter astrometric solutions. The values will be used to ascertain the variability of the sample.

\begin{figure*}
 \centerline{\includegraphics*[width=\linewidth]{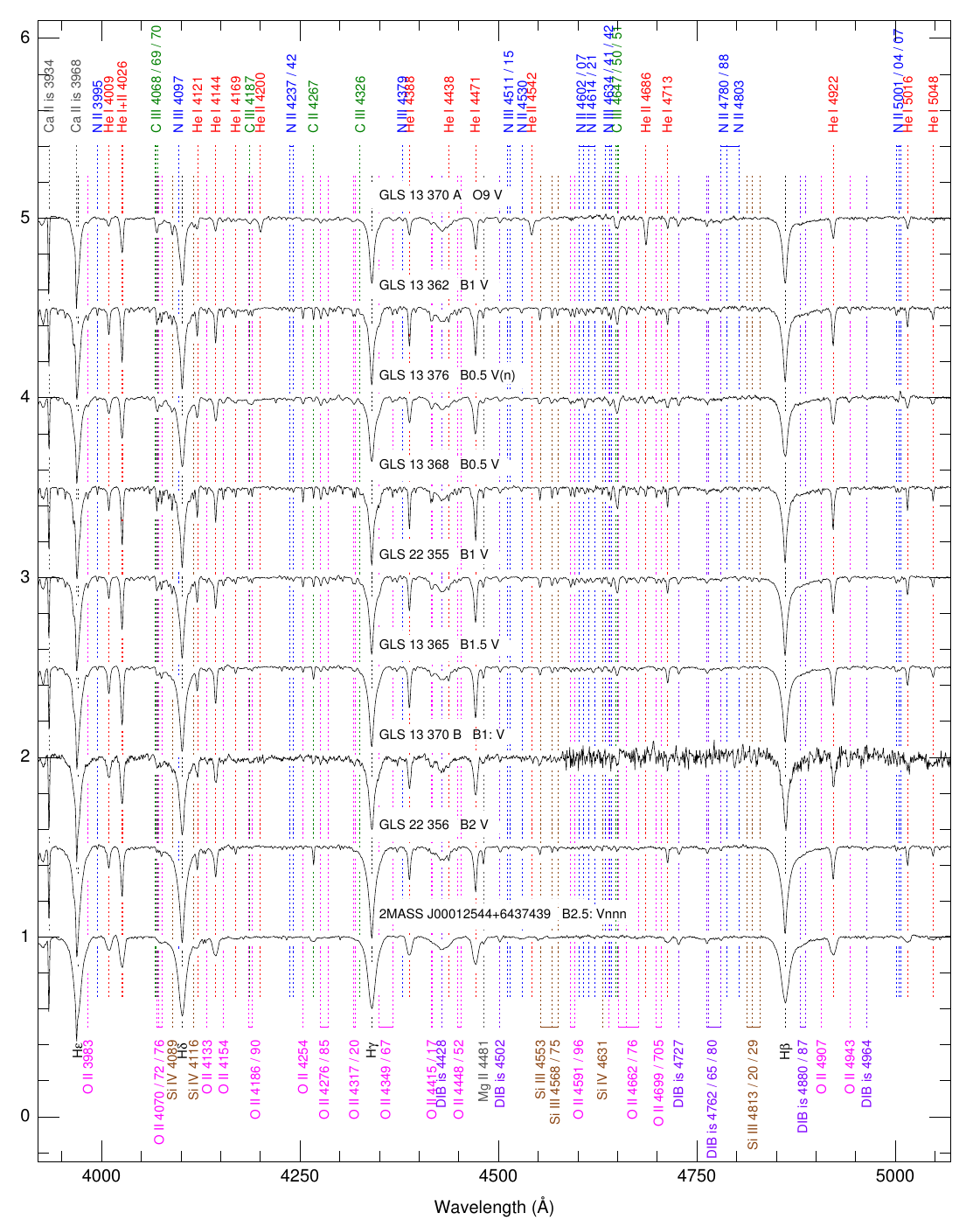}}
 \caption{ALS spectrograms for the nine Stock~18 stars observed with OSIRIS/GTC.}
 \label{ALS_spectra}   
\end{figure*}

\section{The cluster core}

$\,\!$\indent The left panel of Fig.~\ref{Stock_18_core} shows the $1.2\arcmin\times1.2\arcmin$ region around $\alpha = 0.400^\circ$, $\delta = 62.626^\circ$ (or 00:01:36, $+$64:37:33.6). Given the relatively small number of cluster members, the exact position of the center is defined only within 2\arcsec\ in RA or in declination. The value used here was determined by fitting a King profile, as explained below, and is consistent with the literature values in Table.~\ref{Stock18_results}. The right panel of Fig.~\ref{Stock_18_core} shows our AstraLux image of the core centered around the GLS~\num{13370} system with the WDS \citep{Masoetal01} components indicated.

Figure~\ref{Stock_18_core} includes cluster members and non-members alike. To help differentiate between them, the represented proper motions have the mean cluster proper motion (derived later) subtracted. Most of the objects shown have small arrows, indicating that they are likely cluster members. The brightest stars have small proper motions relative to the cluster and most of the objects with large arrows are faint, as expected for a cluster-dominated population at the core. The core is rich in very red stars that correspond to a PMS population, as we will later see. 

\begin{table*}
\caption{Stars in Stock~18 with ALS spectral classifications sorted by \GGGc. The first spectral classifications are from this paper and the literature ones are from \citet{russeil2007revised}.}
\label{spclas}
\centerline{
\begin{tabular}{lllllrl}
\hline
Name                      & ST    & LC & qual. & \mci{GOS/GBS ID}     & \mci{\textit{Gaia} DR3 ID} & SC lit. \\
\hline
GLS~\num{13370}~A         & O9    & V  & ---   & 117.62$+$02.27$\_$01 & \num{431769183925254272}   & O9~V    \\
GLS~\num{13362}           & B1    & V  & ---   & 117.57$+$02.28$\_$01 & \num{431771863984836352}   & B3~V    \\
GLS~\num{13376}           & B0.5  & V  & (n)   & 117.64$+$02.23$\_$01 & \num{431768943407111424}   & B1~V    \\
GLS~\num{13368}           & B0.5  & V  & ---   & 117.60$+$02.16$\_$01 & \num{431766950542328448}   & ---     \\
GLS~\num{22355}           & B1    & V  & ---   & 117.63$+$02.32$\_$01 & \num{431775334318395904}   & ---     \\
GLS~\num{13365}           & B1.5  & V  & ---   & 117.58$+$02.24$\_$01 & \num{431771623466692608}   & B2~V    \\
GLS~\num{13370}~B         & B1:   & V  & ---   & 117.62$+$02.27$\_$02 & \num{431769183920113024}   & ---     \\
GLS~\num{22356}           & B2    & V  & ---   & 117.63$+$02.27$\_$01 & \num{431769287004467584}   & ---     \\
2MASS~J00012544$+$6437439 & B2.5: & V  & nnn   & 117.60$+$02.28$\_$01 & \num{431769252644723456}   & ---     \\
\hline
\end{tabular}
}
\end{table*}

\begin{table*}
\caption{\textit{Gaia}~DR3 information for the stars in Table~\ref{spclas}. F stands for Final sample.}
\label{Gaia_stars}
\centerline{
\begin{tabular}{lccr@{$\pm$}lcr@{$\pm$}lc}
\hline
Name                      & \GGGc   & \BPRP  & \mcii{\pic}  & RUWE & \mcii{\sG}    & Samples \\
                          & (mag)   & (mag)  & \mcii{(mas)} &      & \mcii{(mmag)} &         \\
\hline
GLS~\num{13370}~A         & 10.0128 & 0.7410 & 0.594&0.125  & 4.73 & 11.8&0.2      & 1+,F   \\
GLS~\num{13362}           & 11.0369 & 0.5027 & 0.352&0.030  & 1.09 &  6.1&0.5      & 1+,F   \\
GLS~\num{13376}           & 11.4609 & 0.5436 & 0.382&0.029  & 0.95 &  5.4&1.5      & 1+,F   \\
GLS~\num{13368}           & 11.5492 & 0.5766 & 0.322&0.062  & 1.49 &  4.2&1.3      & 2+,F   \\
GLS~\num{22355}           & 12.1489 & 0.8819 & 0.367&0.023  & 0.98 &  3.0&0.4      & 1+,F   \\
GLS~\num{13365}           & 12.4640 & 0.6691 & 0.357&0.023  & 0.96 &  1.1&1.0      & 1+,F   \\
GLS~\num{13370}~B         & 12.5150 & ---    & 0.325&0.061  & 1.13 &  \mcii{---}   & F      \\
GLS~\num{22356}           & 12.6082 & 0.8241 & 0.371&0.031  & 1.26 &  2.6&0.1      & 1+,F   \\
2MASS~J00012544$+$6437439 & 12.7663 & 0.8206 & 0.373&0.060  & 1.66 &  1.6&1.1      & 1+,F   \\
\hline
\end{tabular}
}
\end{table*}

\begin{table*}
\caption{Additional bright stars within 10\arcmin\ of the cluster center without ALS spectral classifications}
\label{Gaia_additional}
\centerline{
\begin{tabular}{lrrcl}
\hline
Name                      & \mci{\textit{Gaia} DR3 ID} & \mci{\GGGc} & Samples & \mci{Comments}                                 \\
                          &                            & \mci{(mag)} &         &                                                \\
\hline
BD~$+$63~2093~A           & \num{431769046481178624}   &  9.9637     & none    & Foreground star near the cluster center        \\
BD~$+$63~2093~B           & \num{431769046481178112}   & 11.0123     & none    & Foreground star near the cluster center        \\
2MASS~J00003565$+$6440364 & \num{431773272734081152}   & 11.2302     & 4       & Likely red giant                               \\
2MASS~J00023159$+$6430483 & \num{431767358552864000}   & 12.3012     & 4       & Likely RC star slightly in the foreground      \\
2MASS~J00020363$+$6435079 & \num{431768153133139456}   & 12.5826     & 4       & Likely B star slightly in the foreground       \\
2MASS~J00013526$+$6440324 & \num{431775330015871872}   & 12.6918     & 1+,F    & Likely B star in cluster                       \\
\hline
\end{tabular}
}
\end{table*}

The right panel of Fig.~\ref{Stock_18_core} shows an excellent correspondence with the \textit{Gaia} chart, as evidenced from the separations, position angles, and magnitude differences derived from our PSF fitting shown in Table~\ref{GLS_13_370_pairs}. The most important difference is the first detection ever of a secondary component in GLS~\num{13370}~A, which leads to its decomposition in Aa and Ab, and that is not present among the \textit{Gaia} sources. Given the large magnitude difference, Ab is likely to be an intermediate-mass star that is too faint to influence the spectral classification of the Aa,Ab pair and, for that reason, we keep labelling the system as GLS~\num{13370}~A. There are very small differences between the parameters derived from AstraLux and from \textit{Gaia} in Table~\ref{GLS_13_370_pairs}, so the influence of Ab is likely small in the \textit{Gaia} position. However, GLS~\num{13370}~A has large RUWE and parallax uncertainty \sigmae\ and a somewhat anomalous proper motion in \textit{Gaia}~DR3, which are the likely effect of the hidden Ab component in the processing. The $\Delta z$ and $\Delta \GGGc$ measured are very similar for the pairs that include B, D, and E, as expected for OBA stars of similar extinction. The value is slightly more different for the Aa,C pair, where the masses are different enough for the intrinsic color difference between the two bands to become apparent.

\section{The massive-star population of Stock~18}

$\,\!$\indent We show in Fig.~\ref{ALS_spectra} the ALS spectra for the nine stars previously mentioned and we list their spectral classifications sorted by \GGGc\ in Table~\ref{spclas} and a selection of their \textit{Gaia}~DR3 derived properties in Table~\ref{Gaia_stars}. Of the nine spectral types, the first seven are massive stars, GLS~\num{22356} is a borderline case with a mass $\sim$8~\Msun, and 2MASS~J00012544$+$6437439 is likely an intermediate-mass star (but an interesting one as it is a very fast rotator, possibly the result of a merger).

\textit{GLS~\num[detect-all]{13370}~A,B}. This is the only O-type system in Stock~18 and was classified as O9~V by \citet{mayer1973photometry} and as O9.5~V by \citet{Georetal73}. We have already discussed its multiplicity above including the newly discovered Ab component. We are able to separate the A and B components in our long slit but the Aa,Ab pair is too close for its magnitude difference to be resolved (though, in principle, it should be doable with lucky spectroscopy, see \citealt{Maizetal18b,Maizetal21b}). We assign spectral classifications of O9~V and B1:~V to each component, with the uncertainty in the spectral subtype of the B component being caused by the relatively low S/N of its spectrogram, a likely result of spatial deconvolution noise in the extraction process. If the spectra is extracted considering the system as a single star, the derived spectral type would be O9.2~V, a typical effect in systems with this magnitude difference. On the other hand, Ab is unlikely to be influencing the spectral classification of A enough to have shifted its subtype from O8.5 to O9 as the magnitude difference is considerably larger. Given the magnitude difference, Ab is likely to be a mid-B dwarf.

\textit{GLS~\num[detect-all]{13376}} and \textit{GLS~\num[detect-all]{13368}}. These are the two earliest-type B stars in the cluster, both B0.5~V. The main difference between them is that the first is a moderately fast rotator, hence the (n) suffix, while the second one is a slow rotator. In addition, we point out that GLS~{13368} is the farthest massive star from the center of the cluster, a factor that we analyze below.

\textit{GLS~\num[detect-all]{13362}} and \textit{GLS~\num[detect-all]{22355}}. These two stars have the same spectral classification of B1~V yet they differ in more than one magnitude in \GGGc. As we will see later, the reason is the different amounts of dust in their sightlines, a consequence of the differential extinction in Stock~18.  

\textit{GLS~\num[detect-all]{13365}} and \textit{GLS~\num[detect-all]{22356}}. These two stars are classified as B1.5~V and B2~V, respectively. They are likely massive stars but close to the intermediate-mass limit.

\begin{table*}
\caption{CHORIZOS results for the stars in Stock~18. Uncertainties are the direct output of CHORIZOS except for \MG, for which the distance contribution is included from $d=2.91\pm0.10$~kpc.}
\label{CHORIZOS}
\centerline{\addtolength{\tabcolsep}{-2pt}
\begin{tabular}{lcr@{$\pm$}lr@{$\pm$}lr@{$\pm$}lr@{$\pm$}lr@{$\pm$}lr@{$\pm$}lc}
\hline
Name & \Teff & \mcii{LC} & \mcii{\EBV}  & \mcii{\RV} & \mcii{\AV}   & \mcii{\AG}   & \mcii{\MG}   & \chired \\
     & (kK)  & \mcii{}   & \mcii{(mag)} & \mcii{}    & \mcii{(mag)} & \mcii{(mag)} & \mcii{(mag)} &         \\
\hline
GLS~\num{13370}~A,B      & 32.1 & 3.886&0.066 & 0.8048&0.0349 & 3.150&0.199 & 2.528&0.061 & 2.413&0.061 & $-$4.798&0.085 & 0.81 \\
GLS~\num{13362}          & 25.4 & 4.945&0.014 & 0.5473&0.0066 & 3.339&0.074 & 1.827&0.023 & 1.771&0.022 & $-$3.047&0.078 & 0.64 \\
GLS~\num{13376}          & 27.2 & 5.310&0.014 & 0.5833&0.0065 & 3.338&0.067 & 1.947&0.022 & 1.883&0.021 & $-$2.738&0.077 & 0.98 \\
GLS~\num{13368}          & 27.2 & 5.284&0.015 & 0.5925&0.0066 & 3.514&0.072 & 2.082&0.024 & 2.007&0.022 & $-$2.777&0.078 & 1.03 \\
GLS~\num{22355}          & 25.4 & 5.228&0.013 & 0.8089&0.0071 & 3.188&0.051 & 2.578&0.022 & 2.428&0.021 & $-$2.603&0.077 & 0.83 \\
GLS~\num{13365}          & 23.7 & 5.391&0.011 & 0.6123&0.0066 & 3.803&0.067 & 2.328&0.021 & 2.219&0.020 & $-$2.074&0.077 & 0.76 \\
GLS~\num{22356}          & 22.0 & 5.296&0.013 & 0.7372&0.0071 & 3.281&0.058 & 2.418&0.025 & 2.282&0.023 & $-$1.994&0.078 & 0.92 \\
2MASS~00012544$+$6437439 & 20.4 & 5.190&0.014 & 0.7040&0.0073 & 3.485&0.067 & 2.453&0.026 & 2.308&0.024 & $-$1.864&0.078 & 0.36 \\
\hline
\end{tabular}
\addtolength{\tabcolsep}{2pt}}
\end{table*}

\begin{figure*}
 \centerline{\includegraphics*[width=0.49\linewidth]{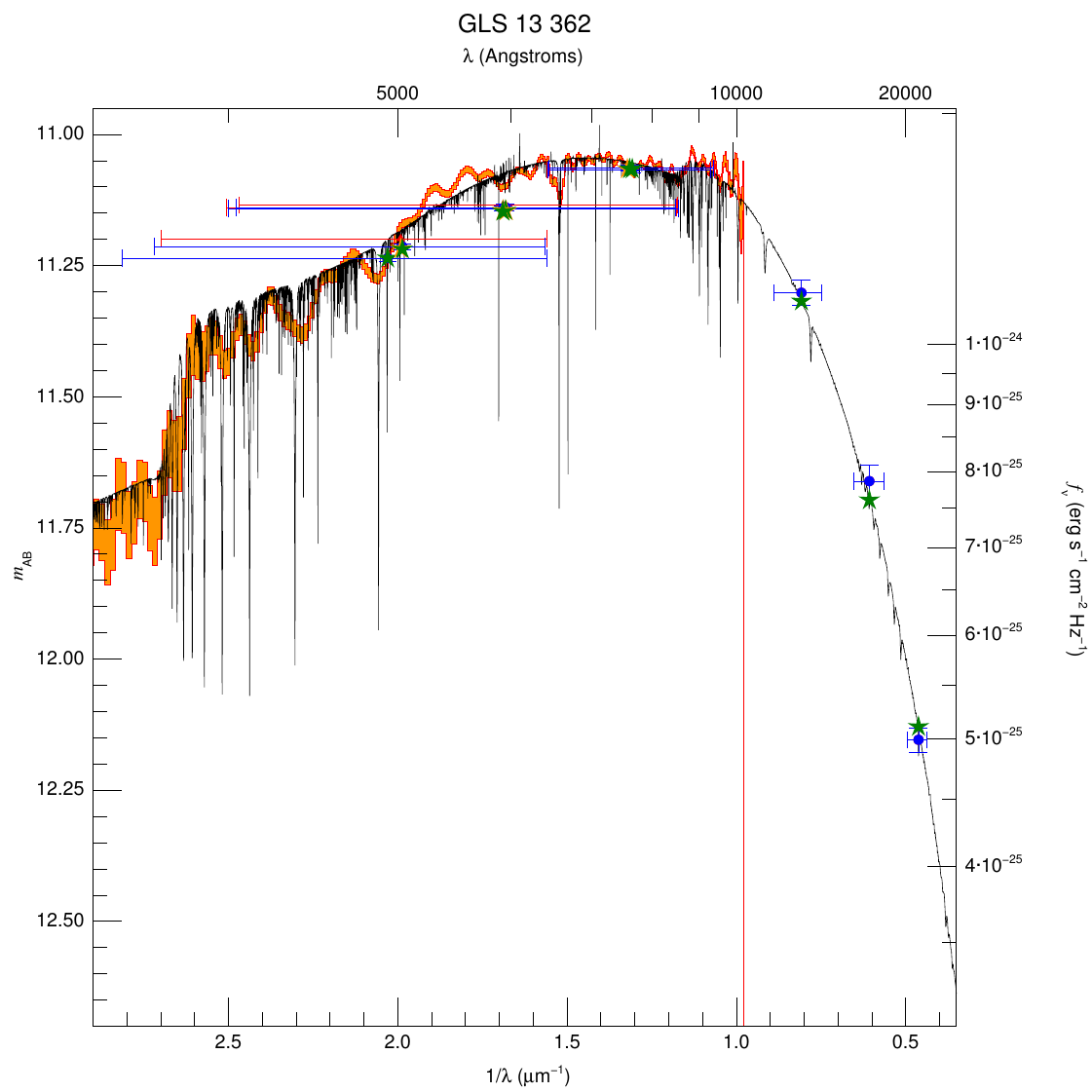} \
             \includegraphics*[width=0.49\linewidth]{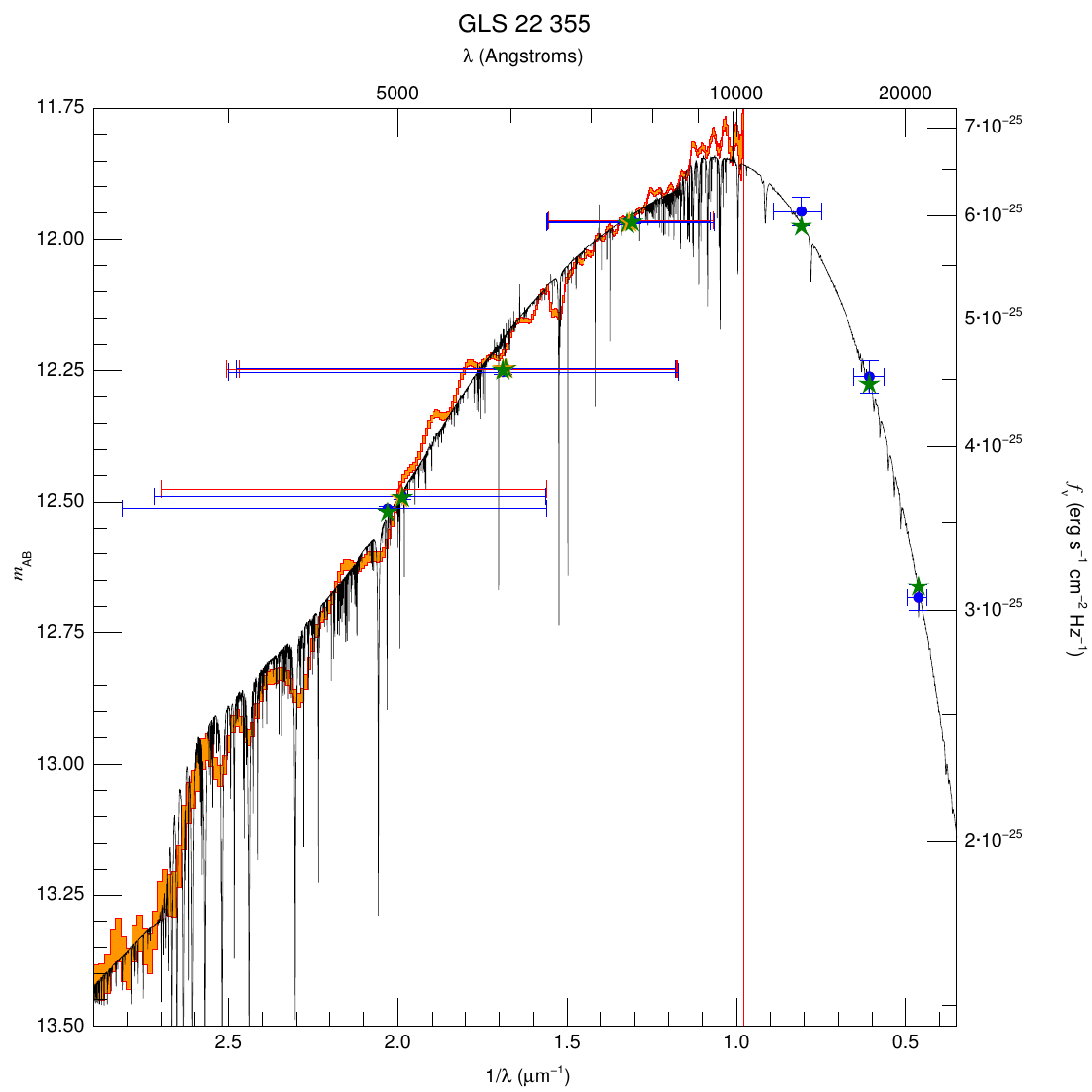}}
 \caption{CHORIZOS SED fits (black lines) to the observed photometry (blue error bars) for the
          least (left panel) and most (right panel) extinguished stars in the CHORIZOS sample.
          The green stars show the model SED magnitudes for each band. The orange banded region shows the 
          \textit{Gaia}~DR3 XP spectrophotometry with errors but note that it is not used for the
          SED fits. The vertical axis spans 1.75 magnitudes in both cases to highlight the significant
          differences in extinction within Stock~18. Note that both stars have the same spectral type
          and, hence, very similar intrinsic SEDs.}
 \label{SEDs}   
\end{figure*}

In addition to those, we also briefly discuss other bright stars in the field. BD~$+$63~2093~A,B is a pair of foreground stars. 2MASS~J00023159$+$6430483 is likely a field red giant at a distance similar to that of Stock~18. 2MASS~J00020363$+$6435079 appears to be a B star close to the distance of Stock~18 but likely in the foreground. Finally, 2MASS J00013526$+$6440324 is a cluster member that, given its \GGGc\ and \BPRP, is likely to be a B star close to the massive-intermediate mass limit.

\section{Extinction analysis}

$\,\!$\indent The results of the CHORIZOS analysis for the stars with ALS spectral types (merging GLS~\num{13370}~A,B, as mentioned above) are given in Table~\ref{CHORIZOS}. The columns give, in order, the star name, the fixed parameter (\Teff, the distance is also set at 2.91~kpc), the three fitted parameters, two extinction-derived parameters, \MG, and the reduced $\chi^2$ (\chired) of the fit. In Figure~\ref{SEDs} we show two examples of the fits.

The fits have \chired\ values close to 1 or even lower, indicating that the magnitude uncertainties are correctly estimated, the filters are well calibrated, the model SEDs are a reasonable representation of the stars, and the family of extinction laws used appropriately describes the phenomenon \citep{MaizBarb18,Maizetal21a}. The minor differences between the fitted SEDs and the observed XP spectrophotometry (Fig.~\ref{SEDs}, note that the spectrophotometry is not used for the fits) are caused by spectral resolution effects and systematic errors in the XP processing \citep{Weiletal20,Weiletal23}.

\begin{figure*}
 \centerline{\includegraphics*[width=0.49\linewidth]{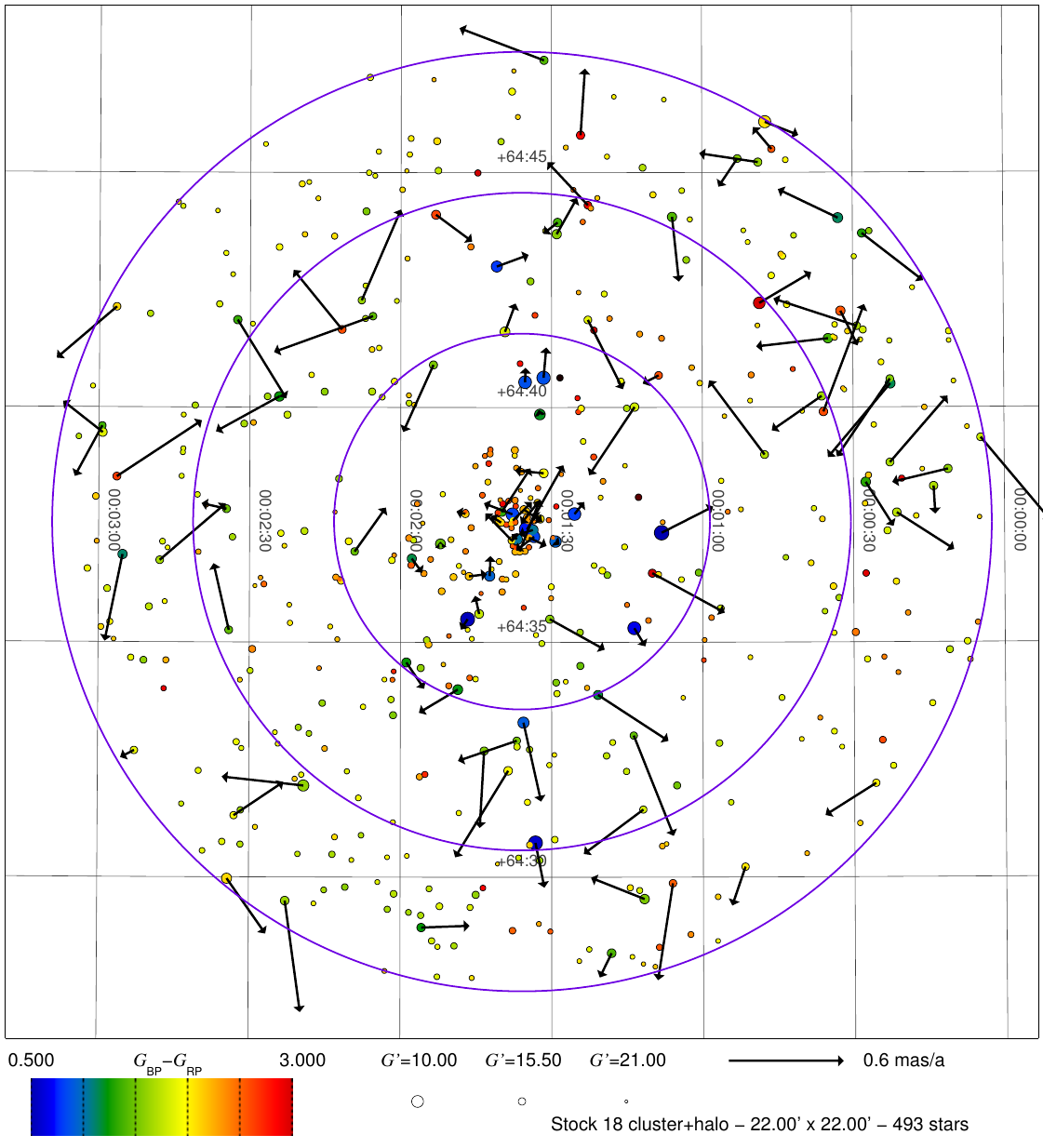} \
             \includegraphics*[width=0.49\linewidth]{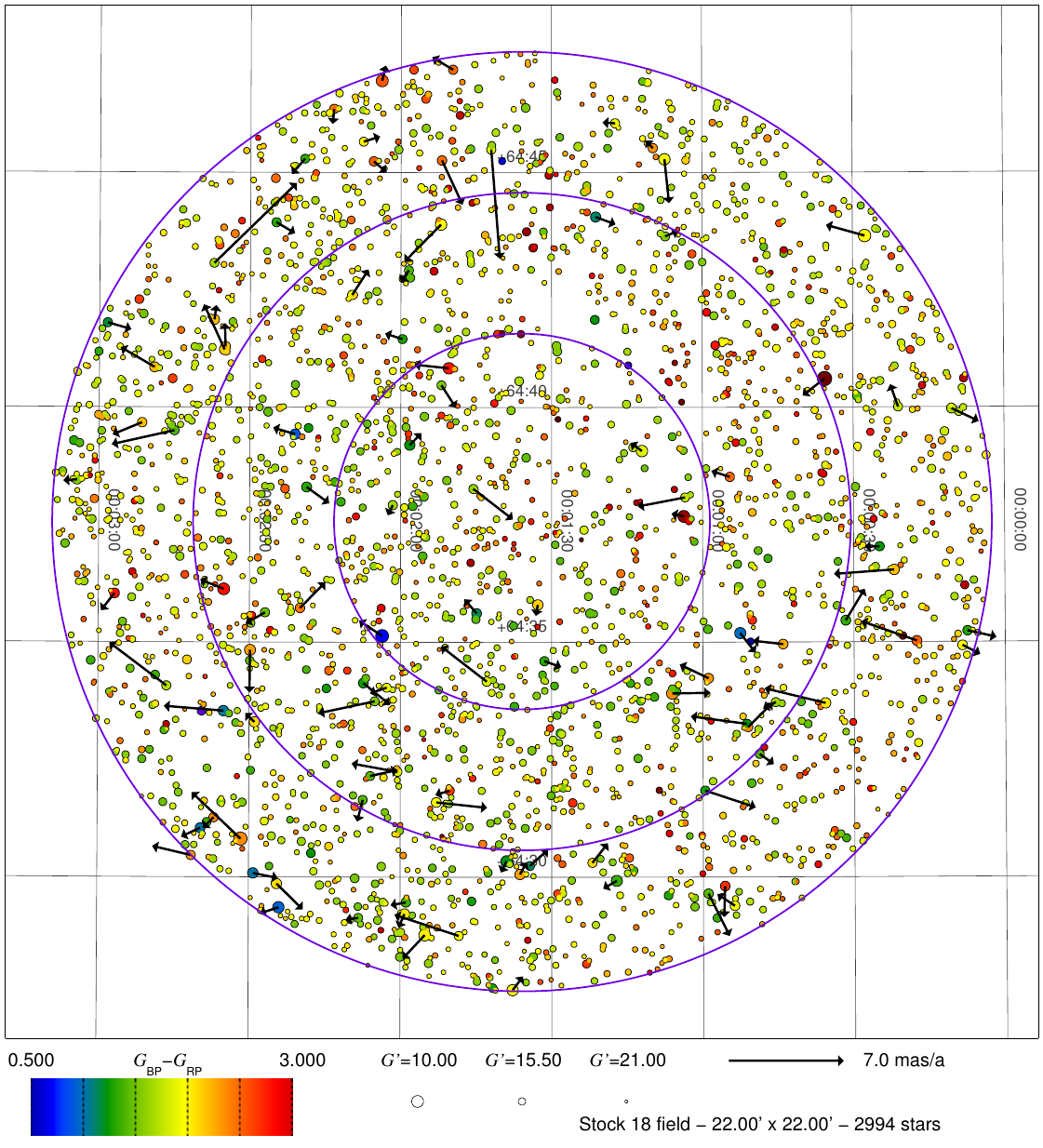}}
 \caption{\textit{Gaia} charts for the stars in samples 1, 2, and 3 (left) and stars
          in differential sample 3-4 (right). The same three circles as in Fig.~\ref{DSS2_WISE}
          are plotted. Symbol size represents \GGGc\ magnitude, symbol color \BPRP, 
          and arrows proper motion. The magnitude and color scales are common to both panels
          (see legend) and stars without a valid \BPRP\ are shown without a color. The proper 
          motions have the mean value derived for sample~1 subtracted and the two scales are 
          different, as indicated in the legend. Only the proper motions for the 100 brightest 
          stars in each panel are shown.}
 \label{Gaia_charts}   
\end{figure*}

In the case of GLS~\num{13370}~A,B we excluded the $K$ photometry from the fit, further (see above) reducing the bands used from six to five, as the magnitude in that band is inconsistent with the rest. The difference implies a $K$-band excess of 0.10~mag, which could be an instrumental effect (as the Aa,Ab,B system is unresolved in 2MASS) or a real physical effect. If it is the latter, it could indicate the existence of an accretion disk in the system (but no obvious H$\alpha$ in emission is seen in the \textit{Gaia} RP spectrophotometry and our ALS spectroscopy does not reach there) or that either Ab or B are subject to a considerably higher extinction than Aa. That does not seem to be the case for GLS~\num{13370}~B, given its spectral type consistent with its \GGGc\ magnitude and the similar $\Delta m$ across bands for the Aa,B pair in Table~\ref{GLS_13_370_pairs}. That leaves Ab as the possible culprit. If that were the case, Ab could be a massive star still embedded in dust, in a situation similar to the companion of Herschel~36 \citep{Gotoetal06,Maizetal15d}. More data in the form of e.g. high-resolution imaging in different IR bands are needed to test that hypothesis.

With the exception of GLS~\num{13362}, all B stars have fitted luminosity classes of 5.2-5.4, with 5.0 being a typical dwarf for that spectral type and 5.5 being a ZAMS object. This is a sign of the young age of the cluster. The higher value of GLS~\num{13362} indicates that it possibly contains a hidden binary component. The CHORIZOS-derived LC of $\sim$3.9 for GLS~\num{13370}~A,B must be partially caused by the presence of three stars in the system but that could not be the whole story given the relatively large $\Delta m$ between components. As the primary has a spectral type slightly inconsistent with that LC (and the late-O dwarf spectral classification does not show luminosity inconsistencies between He/He and Si/He criteria, see \citealt{Walbetal14}) and is unlikely to have progressed along an isolated evolutionary trajectory, we propose as a possibility that it is the result of a merger.

As previously detected by other authors (Table~\ref{Stock18_results}), there is significant differential reddening, with \EBV\ ranging from 0.55 to 0.81. Given the small uncertainties, the effect is clearly physical, as it can be also seen comparing the two panels in Fig.~\ref{SEDs} which correspond to two stars with very similar intrinsic SEDs but quite different extinctions. 

We also detect variations in \RV\ within the cluster (see \citealt{MaizBarb18} for examples in other clusters with associated \HII\ regions) but they are smaller than those in \EBV. Six of the eight stars are within two sigmas of \RV~=~3.38, with GLS~\num{22355} and GLS~\num{13365} being the two outliers. The average \RV\ value is higher than the canonical one of 3.1, as typical of the extinction produced by dust immersed in ionized gas \citep{MaizBarb18}. 

The uncertainties in the monochromatic extinction \AV\ are smaller than those expected by the simple product of \EBV\ and \RV. This is caused by the anticorrelation between \EBV\ and \RV\ described by \citet{MaizBarb18}. The band-integrated extinction \AG\ is always lower than the monochromatic \AV. This happens because the effective wavelength of the \GGG\ filter is larger than 5495~\AA\ but also due to non-linearity effects \citep{Maiz24}, as the ratio of \AV\ to \AG\ increases with \AV. 

\section{Cluster membership and distance}

\begin{table}
\caption{Sample characteristics used in this paper for membership.}
\label{samples}
\centerline{\addtolength{\tabcolsep}{-2pt}
\begin{tabular}{crccrcl}
\hline
Sample & $r$       & \rmu           & & \mciii{All samples}                                \\
\cline{5-7}
       & (\arcmin) & (mas/a)        & & \alphac         & = & \phantom{$-$}$0.400^\circ$   \\
\cline{1-3}
1      &  4        & \phantom{2}0.6 & & \deltac         & = & \hspace{0.5mm}$64.626^\circ$ \\
2      &  7        & \phantom{2}0.6 & & \pmrac          & = & $-$2.68 mas/a                \\
3      & 10        & \phantom{2}0.6 & & \pmdecc         & = & $-$0.66 mas/a                \\
4      & 10        & 20.0           & & \Cstar          & < & \phantom{$-$}0.40            \\
\cline{1-3}
       &           &                & & $\Delta(\BPRP)$ & > & $-$0.50 mag                  \\
       &           &                & & RUWE            & < & \phantom{$-$}5               \\
\cline{5-7}
\end{tabular}
\addtolength{\tabcolsep}{2pt}}
\end{table}

\begin{table*}
\caption{Membership and distance results for the four samples.}
\centerline{
\renewcommand{\arraystretch}{1.2}
\begin{tabular}{lrrrrrr@{$\pm$}lr@{$\pm$}lr@{$\pm$}lr@{}l}
\hline
ID & $N_{*,0}$ & $N_*$ & $t_\varpi$ & $t_{\mu_{\alpha *}}$ & $t_{\mu_{\delta}}$ & \mcii{\pmrag}  & \mcii{\pmdecg} & \mcii{\pig}  & \mcii{$d$}              \\
   &           &       &            &                      &                    & \mcii{(mas/a)} & \mcii{(mas/a)} & \mcii{(mas)} & \mcii{(kpc)}            \\
\hline
1  &       154 &   150 &       0.96 &                 1.46 &               1.49 & $-$2.670&0.024 & $-$0.658&0.024 & 0.344&0.012  & 2.91&$^{+0.10}_{-0.10}$ \\
2  &       287 &   278 &       0.94 &                 1.73 &               1.93 & $-$2.686&0.023 & $-$0.684&0.023 & 0.335&0.011  & 2.99&$^{+0.10}_{-0.10}$ \\
3  &       513 &   493 &       0.94 &                 1.94 &               1.98 & $-$2.667&0.022 & $-$0.706&0.022 & 0.325&0.011  & 3.08&$^{+0.11}_{-0.10}$ \\
4  &      3987 &  3485 &       1.12 &                13.23 &               8.62 & $-$2.208&0.022 & $-$0.684&0.022 & 0.344&0.010  & 2.91&$^{+0.09}_{-0.08}$ \\
\hline
\end{tabular}
\renewcommand{\arraystretch}{1.0}
}
\label{sample_results}                  
\end{table*}

\begin{figure*}
 \centerline{\includegraphics*[width=0.49\linewidth]{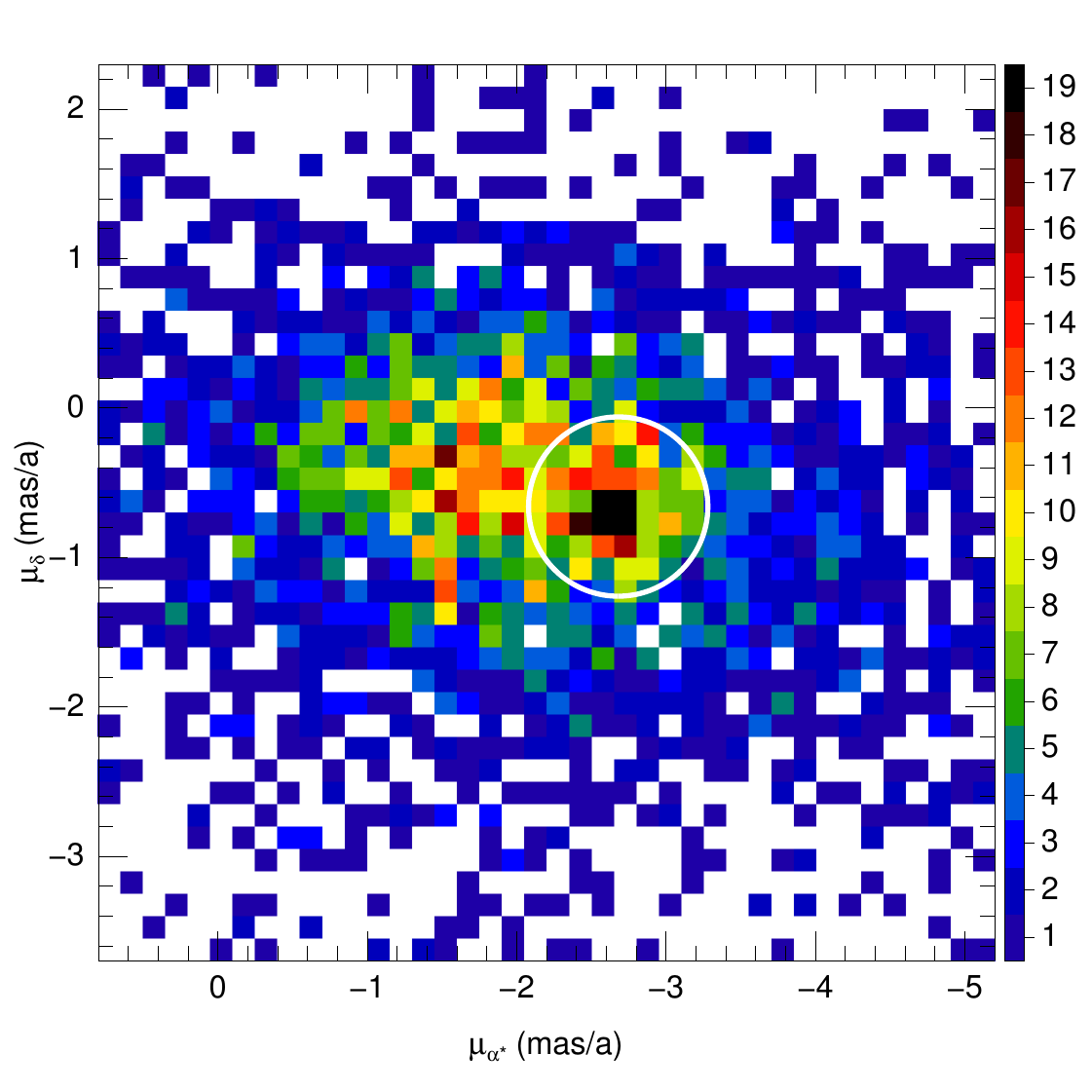} \
             \includegraphics*[width=0.49\linewidth]{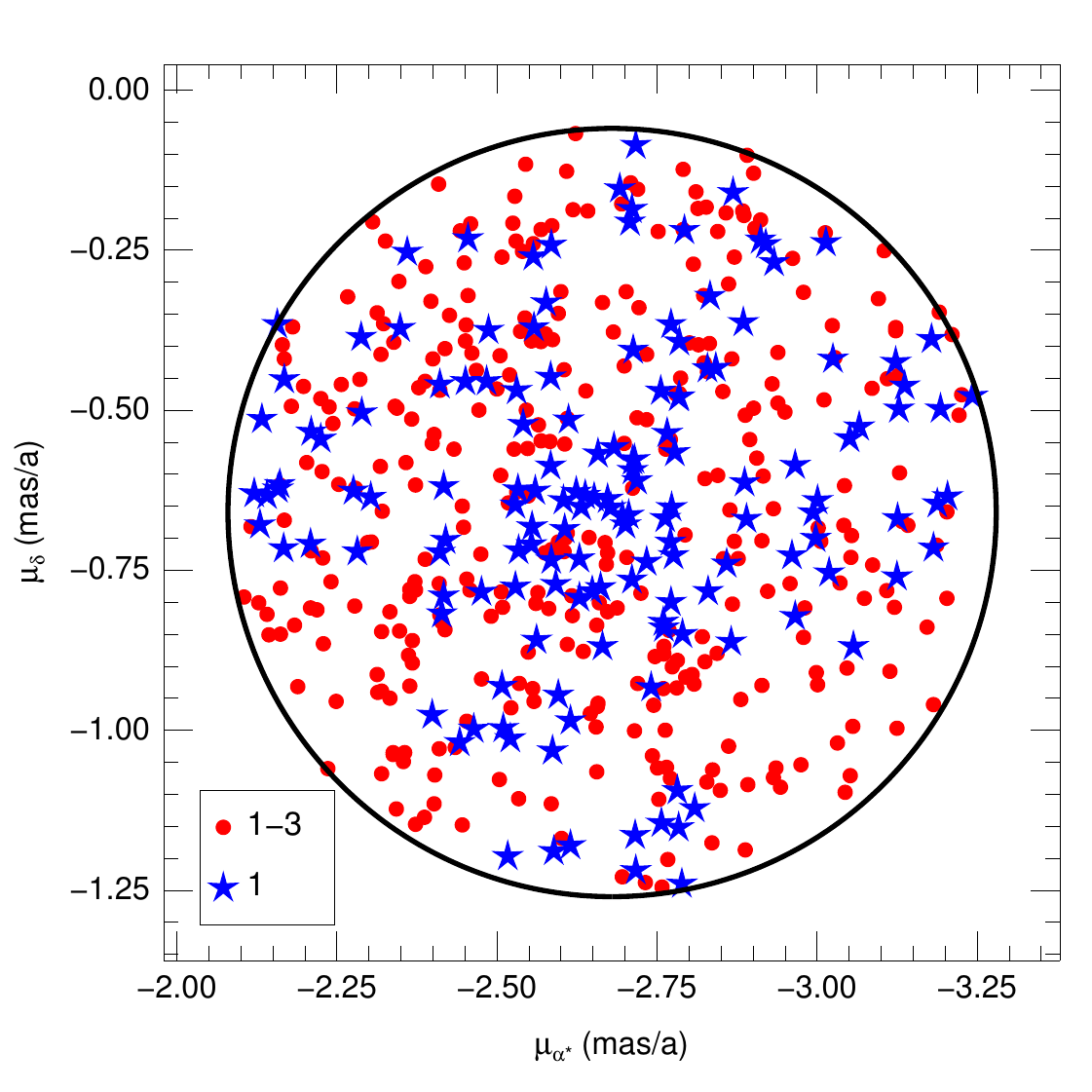}}
 \caption{Proper motion diagrams for different samples in this paper. (left) Histogram for 
          sample 4, excluding the regions with high relative proper motions for visibility
          purposes. The white circle marks the selected region for samples 1 to 3. (right)
          Individual proper motions for sample 1 (blue stars) and for the differential
          1-3 sample (red circles).}
 \label{Gaia_pm}   
\end{figure*}

\begin{figure}
 \centerline{\includegraphics*[width=\linewidth]{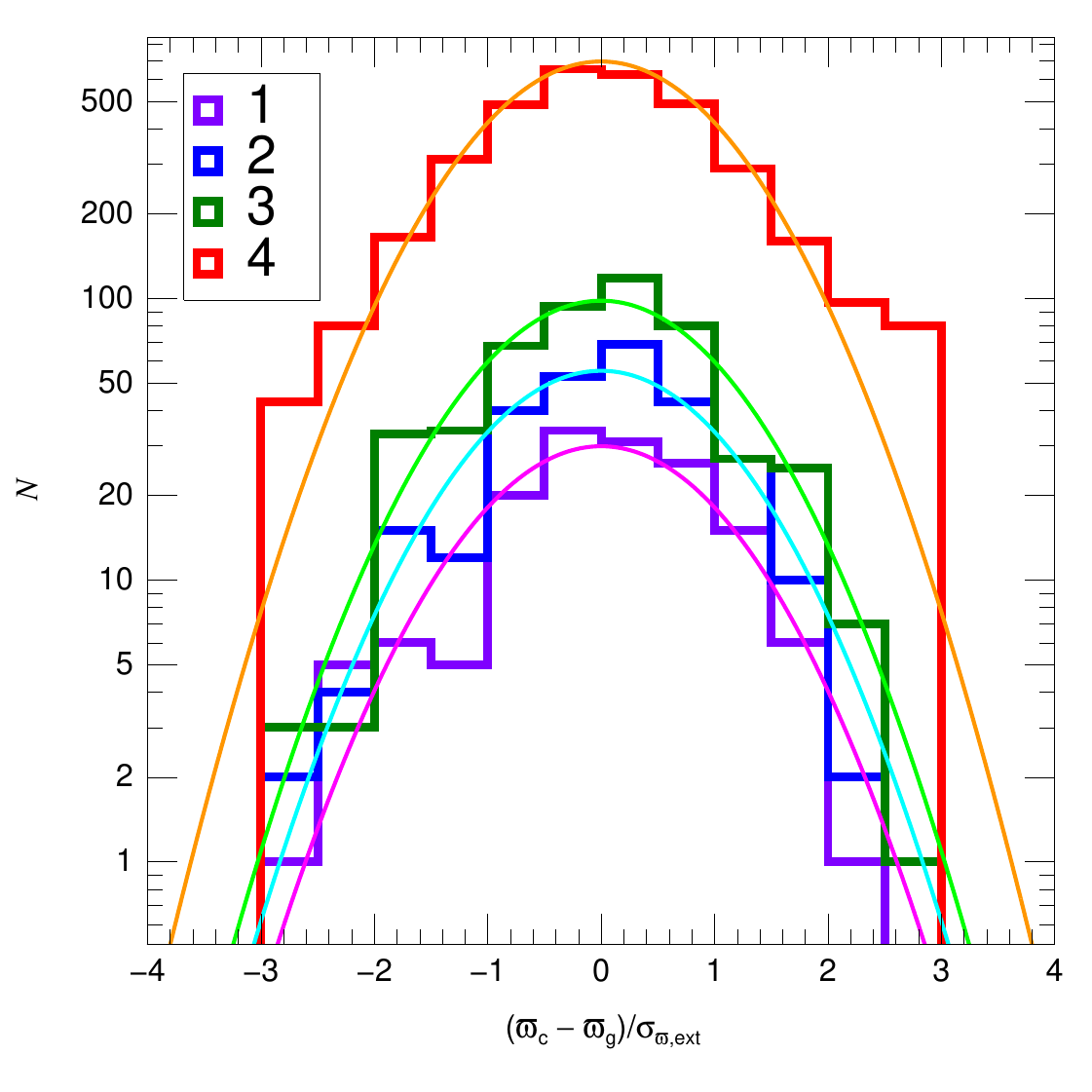}}
 \caption{Normalized parallax histograms for samples 1-4. The smooth curves are the expected distributions.}
 \label{Gaia_normpar}   
\end{figure}

\begin{figure}
 \centerline{\includegraphics*[width=\linewidth]{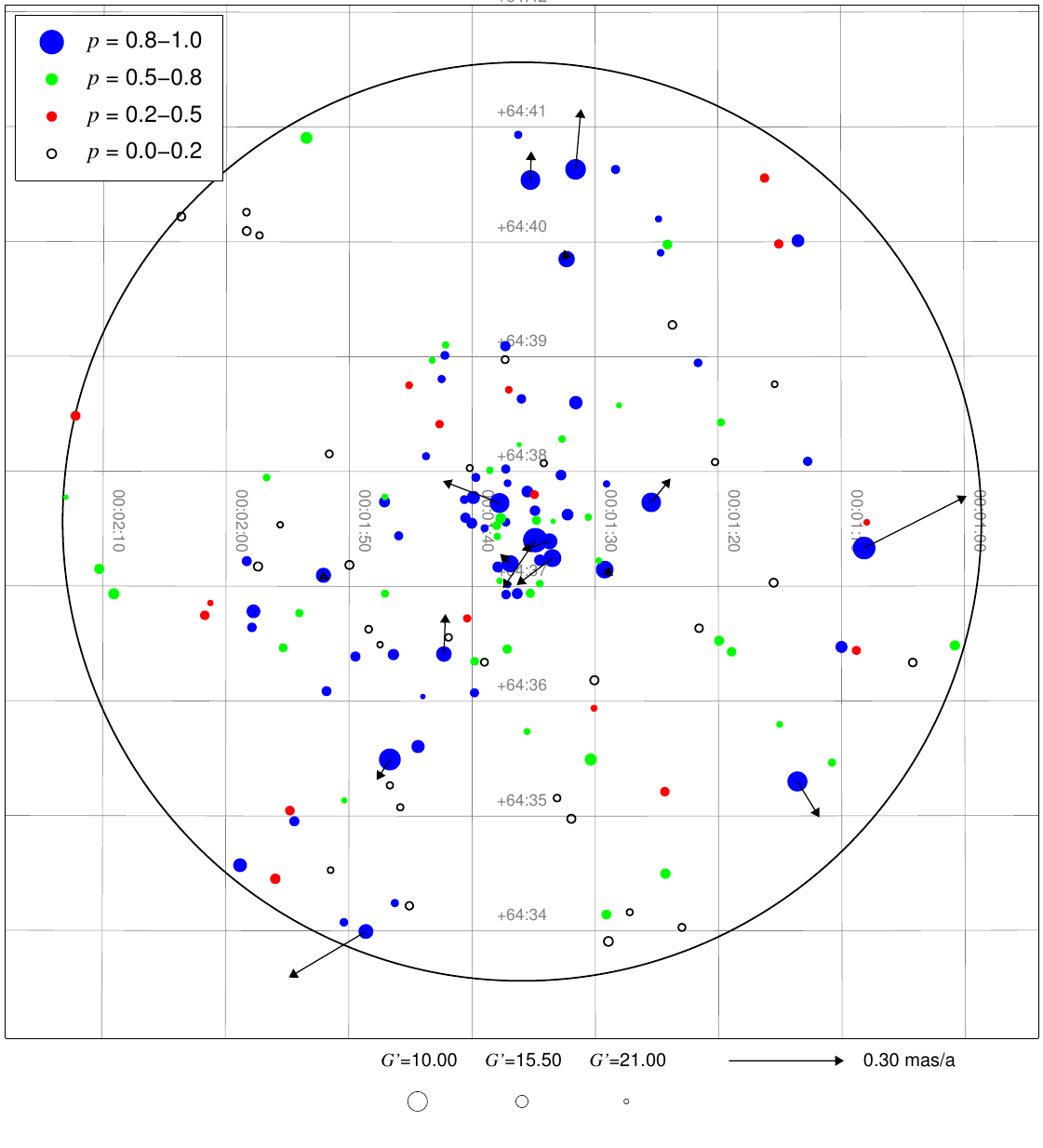}}
 \caption{\textit{Gaia} $9\arcmin\times9\arcmin$ chart for sample 1 with symbol size used to represent 
          magnitude and a coloring scheme that reflects the probability of belonging to the cluster.
          Arrows represent the proper motion with respect to the sample average, with the value for 
          GLS~\num{13370}~B substituting that of GLS~\num{13370}~A, for the 16 brightest stars in 
          sample 1.}
 \label{Gaia_chart_4}   
\end{figure}

\subsection{Sample definitions}

$\,\!$\indent To select the cluster membership and calculate the distance to Stock~18 we use four primary samples (1 to 4) selected using the procedure described above and the criteria listed in Table~\ref{samples}. All samples have the same center, with the radius increasing from 4\arcmin\ in sample~1 to 7\arcmin\ in sample~2 and 10\arcmin\ in sample~3 (Fig~\ref{DSS2_WISE}). $r$ is maintained at 10\arcmin\ for sample~4 but in that case \rmu\ is increased from 0.6~mas/a to 20~mas/a. The values listed in Table~\ref{samples} were selected iteratively, as with other groups in the Villafranca project, with the idea of using sample~1 for the cluster itself based on the spatial distribution, proper motions, and extinctions of its massive stars. \alphac\ and \deltac\ were determined by fitting King profiles and minimizing the cluster radius, as detailed below.
Samples~2 and 3 are spatially extended versions of sample~1 to determine the extension of a potential halo (or association) around Stock~18. Sample~4 represents the bulk of the Galactic population in \textit{Gaia}~DR3\footnote{The bulk of the Galactic population in a given region of the sky in \textit{Gaia}~DR3 is at a distance determined by competing effects in two directions: on the one hand, the ever increasing volume at a given distance due to the $d^2$ factor and, on the other hand, the dimming effect of distance and extinction that moves stars past the detection limit. Variations in stellar density can play in both directions but most frequently in the second one, especially at a Galactic latitude of more than $2^\circ$ and away from the bulge such as in this case.}, as setting such a large \rmu\ and leaving \Cstar,  $\Delta(\BPRP)$, and RUWE only loosely constrained leave the selection algorithm to operate almost only through the normalized parallax loop. Given the Matrioshka nature of the four samples, with each subsequent one enclosing the previous samples in parameter space\footnote{With only two exceptions, a star in a given sample is also included in the superior ones, hence the notation 2+ in Tables~\ref{Gaia_stars}~and~\ref{Gaia_additional} indicates samples 2, 3, and 4. The two exceptions are two stars in sample 3 not present in sample 4 after being eliminated by the normalized parallax criterion.}, we also define differential samples X-Y as those in sample Y but not in sample X. For example, the differential sample 3-4 refers to stars in the bulk of the Galactic population excluding those within the range of proper motions of the cluster. Later on we also define a fifth sample which we call the final one.

\subsection{Sample results}

$\,\!$\indent The positions, magnitudes, colors, and (a selection of the) proper motions are plotted in Fig.~\ref{Gaia_charts}, proper motion diagrams are given in Fig.~\ref{Gaia_pm}, normalized parallax histograms are shown in Fig.~\ref{Gaia_normpar}, and CAMDs are plotted in Fig~\ref{Gaia_CAMDs}. A summary of the results for samples 1 to 4 is listed in Table~\ref{sample_results}.

All four samples yield very similar distances, all within one sigma of 3.0~kpc. This indicates that the cluster is located at about the same distance (within the uncertainties) as the bulk of the Galactic population. In reality, the stars in the cluster should be located at a small range of distances of a few pc while the bulk of the Galactic population should extend over hundreds of parsecs. However, the external parallax uncertainties in \textit{Gaia}~DR3 are too large to differentiate between the two populations using parallaxes alone, especially as the bulk of the Galactic population is faint (Fig.~\ref{Gaia_CAMDs}) and, hence, their uncertainties are larger than for most of the bright stars in the cluster. 

The normalized parallax histograms for the four samples (Fig.~\ref{Gaia_normpar}) are reasonably well approximated by Gaussians of zero mean and a dispersion of one, indicating that they are consistent with populations at a similar distance and that the external uncertainties are correctly estimated. The values of $t_\varpi$ close to 1.0, as a numerical expression of the above, indicate the same. Note that the normalized parallax criterion eliminates objects with values below $-$3 or above $+$3. There are not enough stars in the first three samples for this criterion to eliminate more than maybe one member of the population, but for sample 4 it is possible that $\sim$10 have been removed (a small fraction of the total, in any case). 

As the cluster and the bulk of the Galactic population are at similar distances, the primary means of differentiating them must be through proper motions. As shown in Table~\ref{sample_results}, samples 1 to 3 have similar average proper motions in the two coordinates (not unexpected given the selection criteria) while those three samples have similar average values of \pmdecg\ when compared to sample~4 but quite different values of \pmra. The effect is seen in the left panel of Fig~\ref{Gaia_pm}, with Stock~18 producing a significant peak in the broader proper motion distribution of sample~4. However, it is clear from that panel that sample~4 significantly contaminates at least sample 3 and quite possibly samples~1~and~2, an issue we analyze next.

\subsection{Contamination and final sample definition}

\begin{figure*}
 \centerline{\includegraphics*[width=0.49\linewidth]{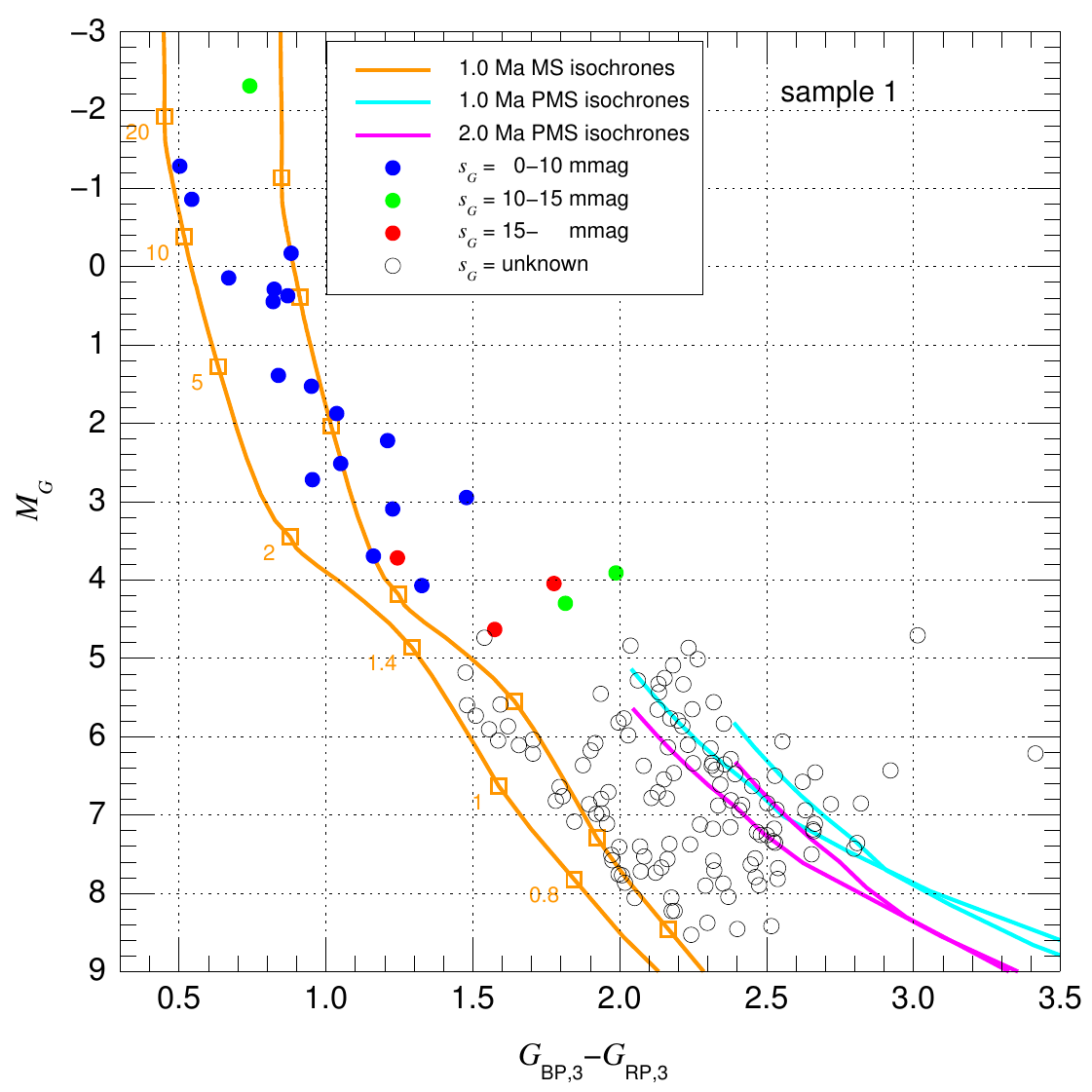} \
             \includegraphics*[width=0.49\linewidth]{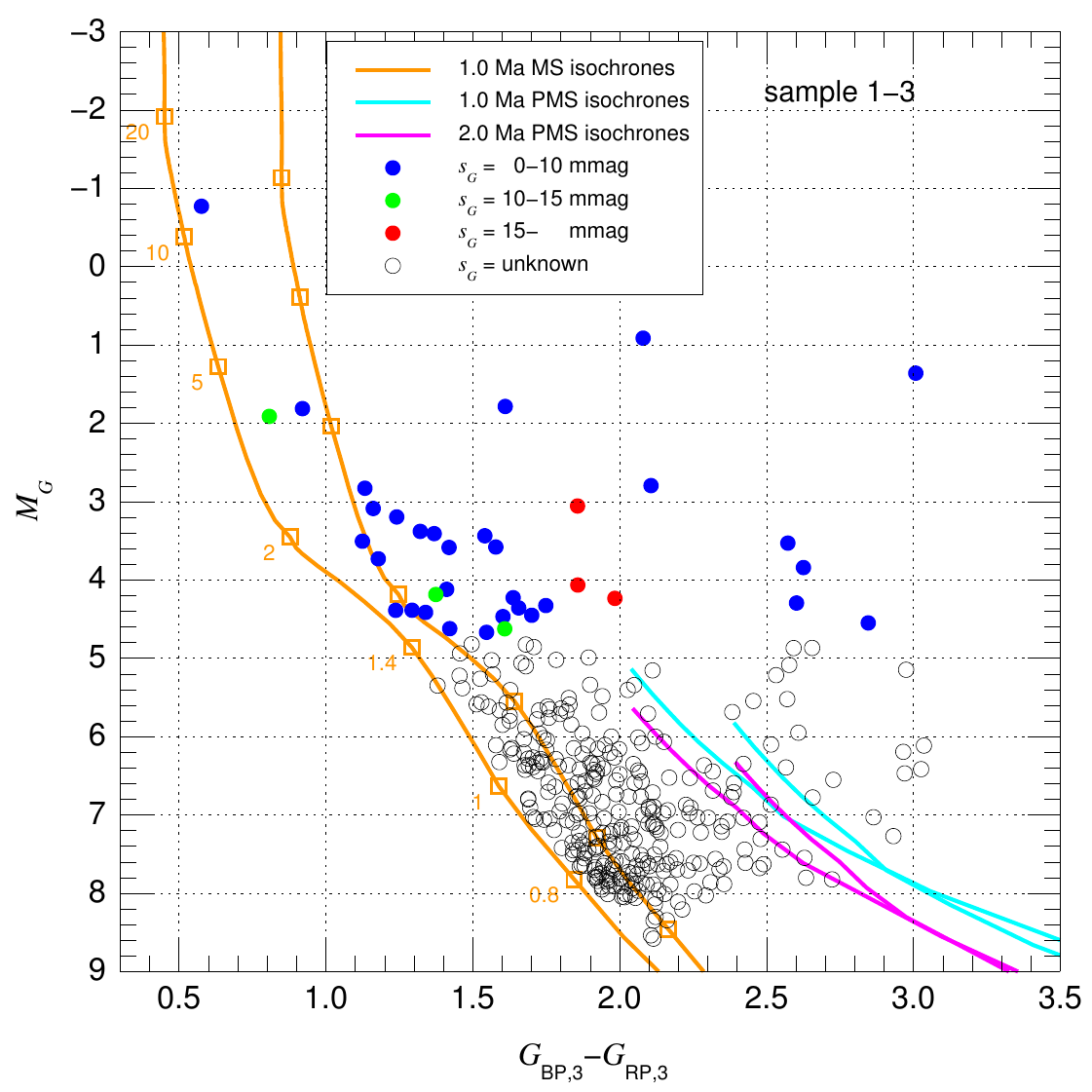}}
 \centerline{\includegraphics*[width=0.49\linewidth]{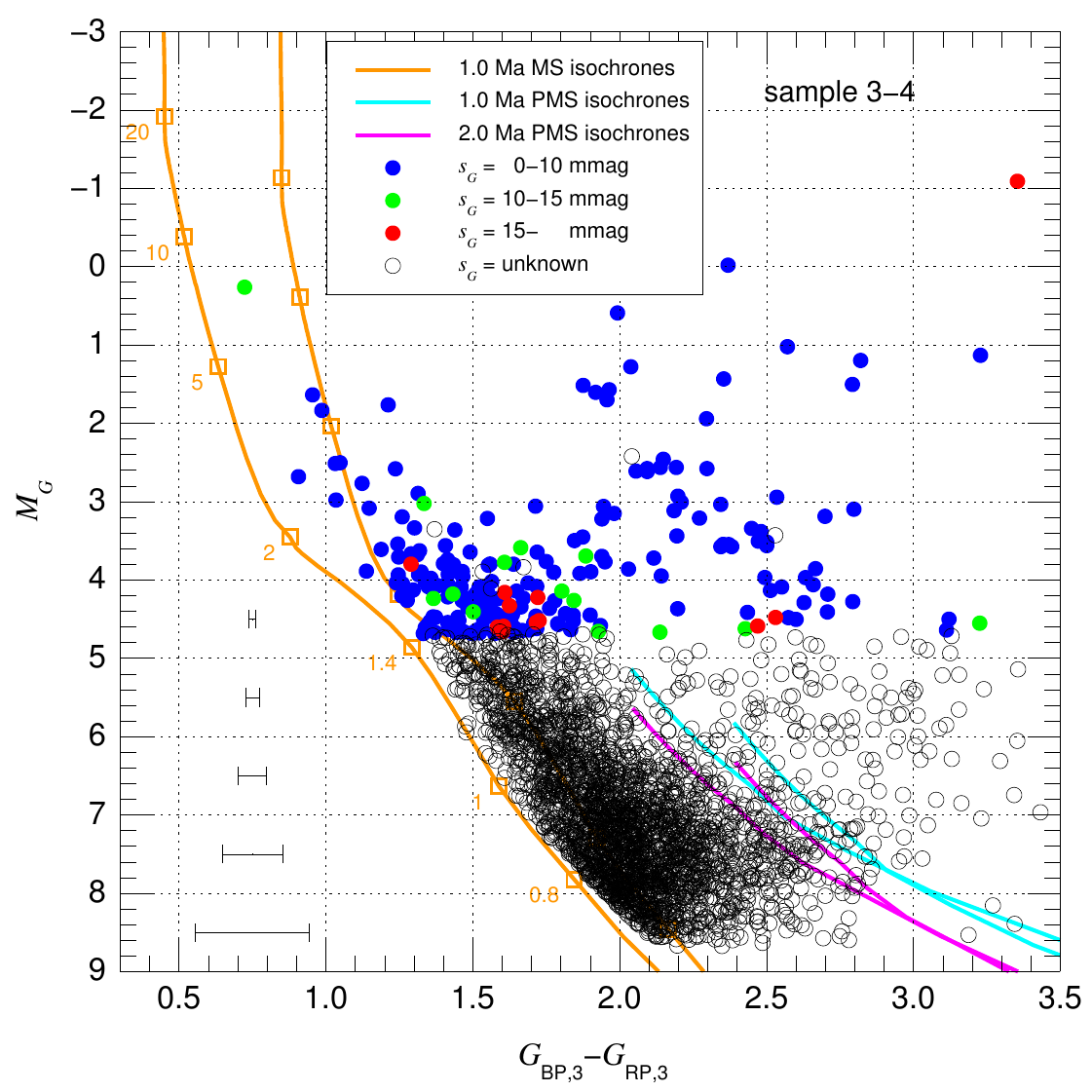} \
             \includegraphics*[width=0.49\linewidth]{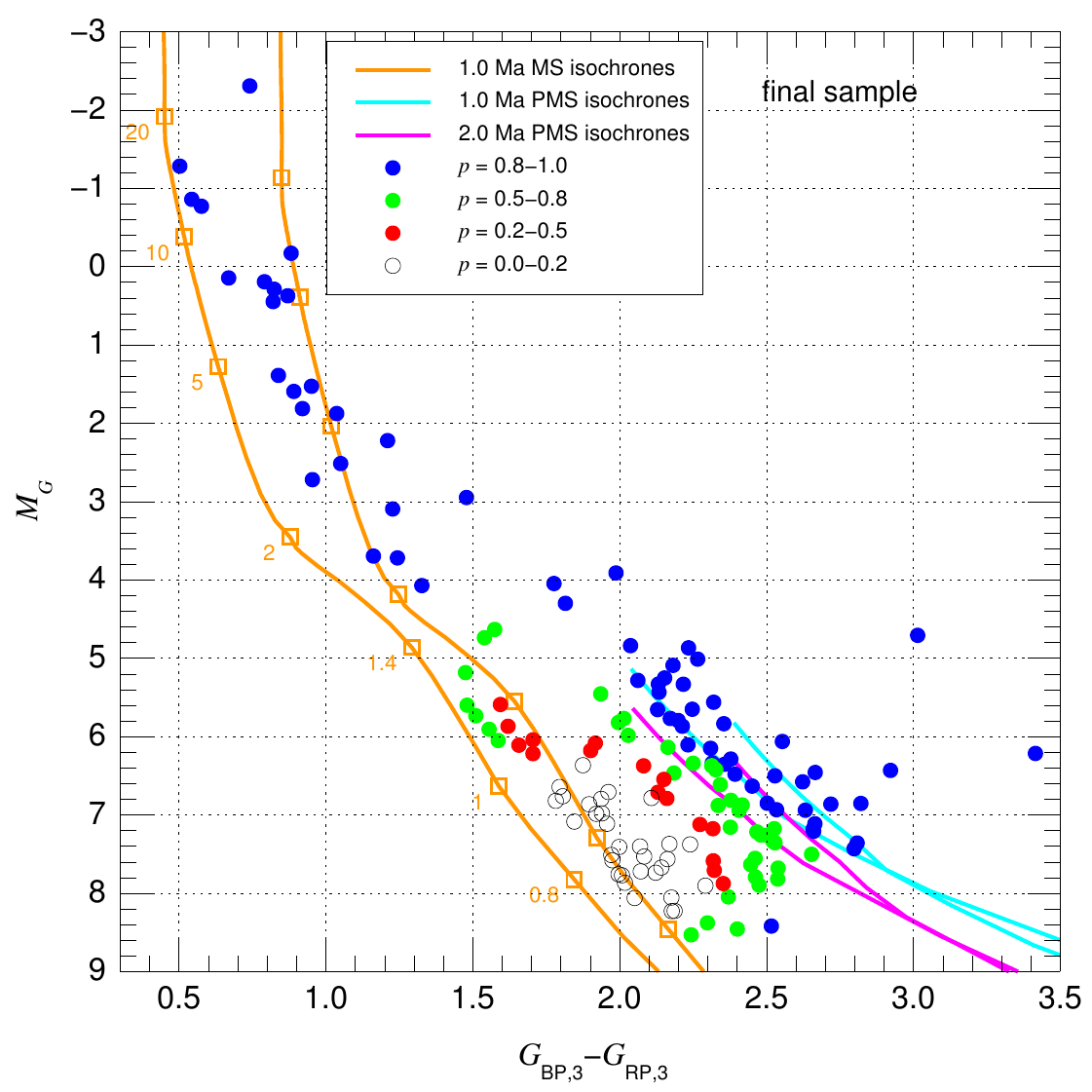}}
 \caption{\textit{Gaia} CAMD for different samples in this paper: (top left) sample 1, (top right)
          differential 1-3 sample, (bottom left) differential 3-4 sample, and (bottom right) final
          sample. All panels include the solar metallicity Geneva-Padova 1.0~Ma MS isochrone of 
          \citet{Maiz13a} and the 1.0~and~2.0~Ma PMS isochrones of \citet{Baraetal15} with the two
          extinctions that span the range measured for Stock~18, \EBV\ of 0.55~and~0.81~mag, in both
          cases with $\RV = 3.38$. Empty squares mark the initial masses in the MS isochrone and the
          PMS isochrones extend to 1.4~\Msun. In the first three panels each sample member is colored 
          according to \sG\ from \citet{Maizetal23}. In the bottom left panel typical color error bars 
          as a function of \MG\ are shown. In the bottom right panel the coloring scheme
          is the same as in Fig.~\ref{Gaia_chart_4}
          and reflects the probability of belonging to the cluster. A distance of 2.91~kpc is assumed for all
          stars (distance modulus of 12.32~mag).}
 \label{Gaia_CAMDs}   
\end{figure*}

\begin{figure*}
 \centerline{\includegraphics*[width=0.5\linewidth]{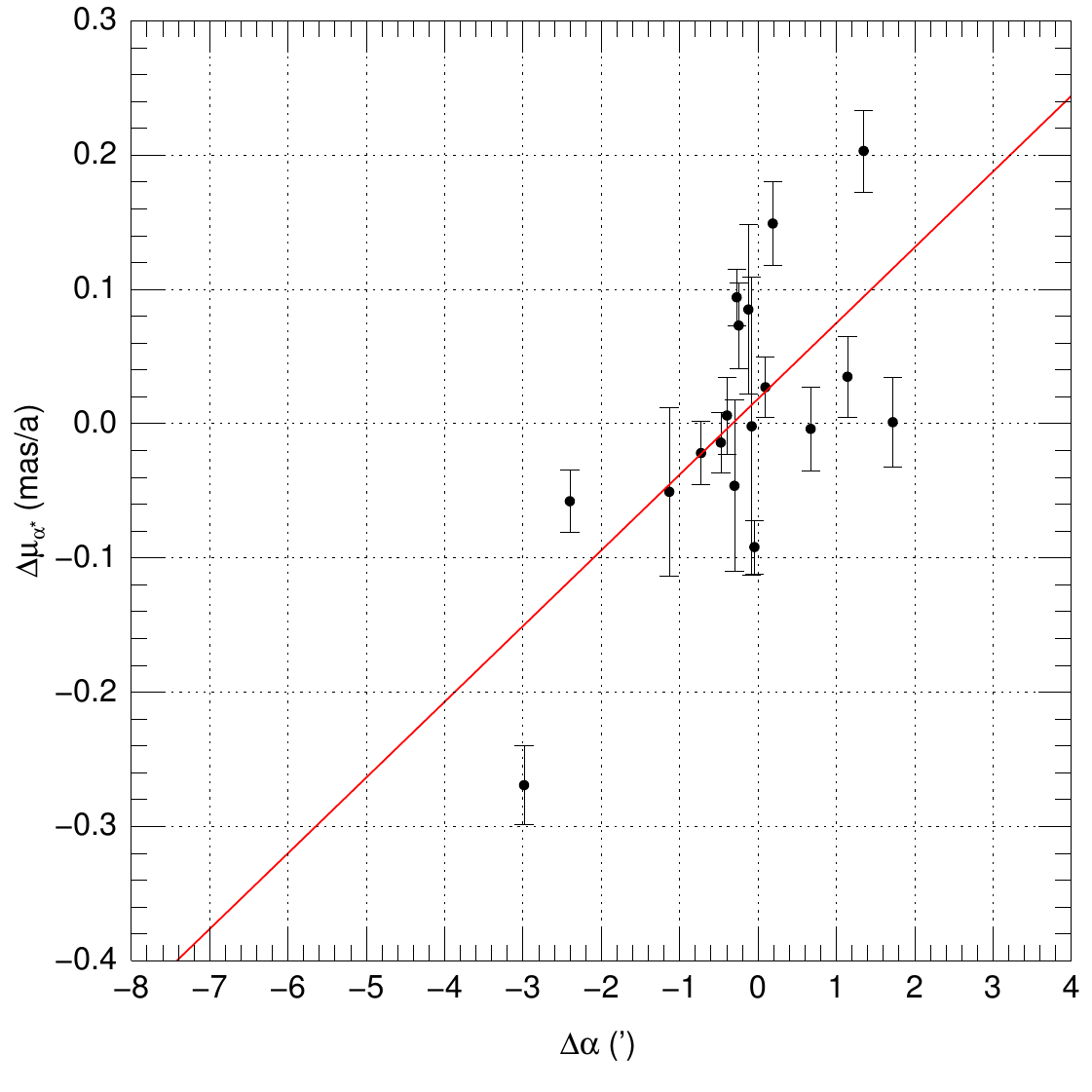} \
             \includegraphics*[width=0.5\linewidth]{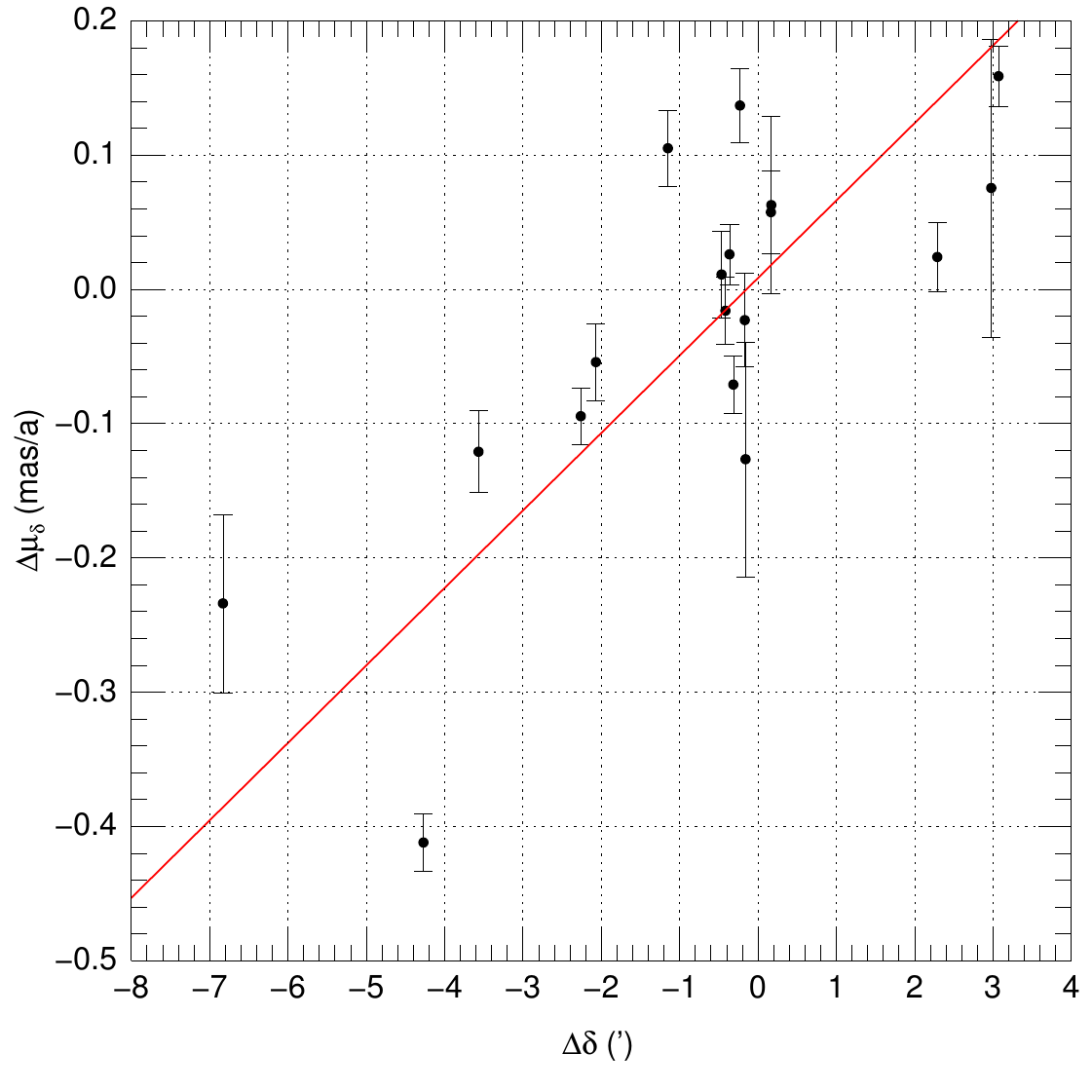}}
 \caption{Right ascension (left) and declination (right) coordinate vs. proper motion for the 16 stars 
          with drawn proper motions in Fig.~\ref{Gaia_chart_4} plus the two brightest stars in the 
          differential 1-2 sample. Error bars show the external uncertainties without the systematic 
          uncertainty, as we are dealing with relative proper motions in a small region of the sky. 
          The red lines are the linear regression fits.}
 \label{Gaia_pm2}   
\end{figure*}

\begin{figure*}
 \centerline{\includegraphics*[width=0.115\linewidth]{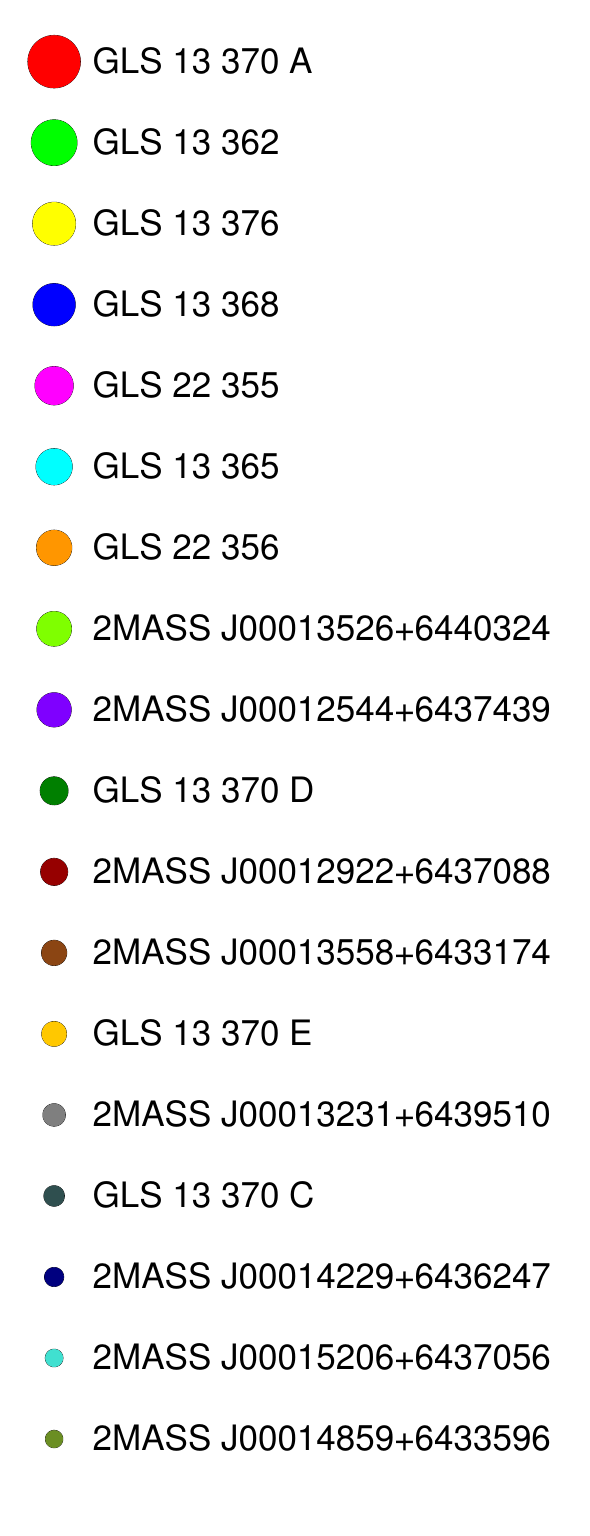}$\!$
             \includegraphics*[width=0.300\linewidth]{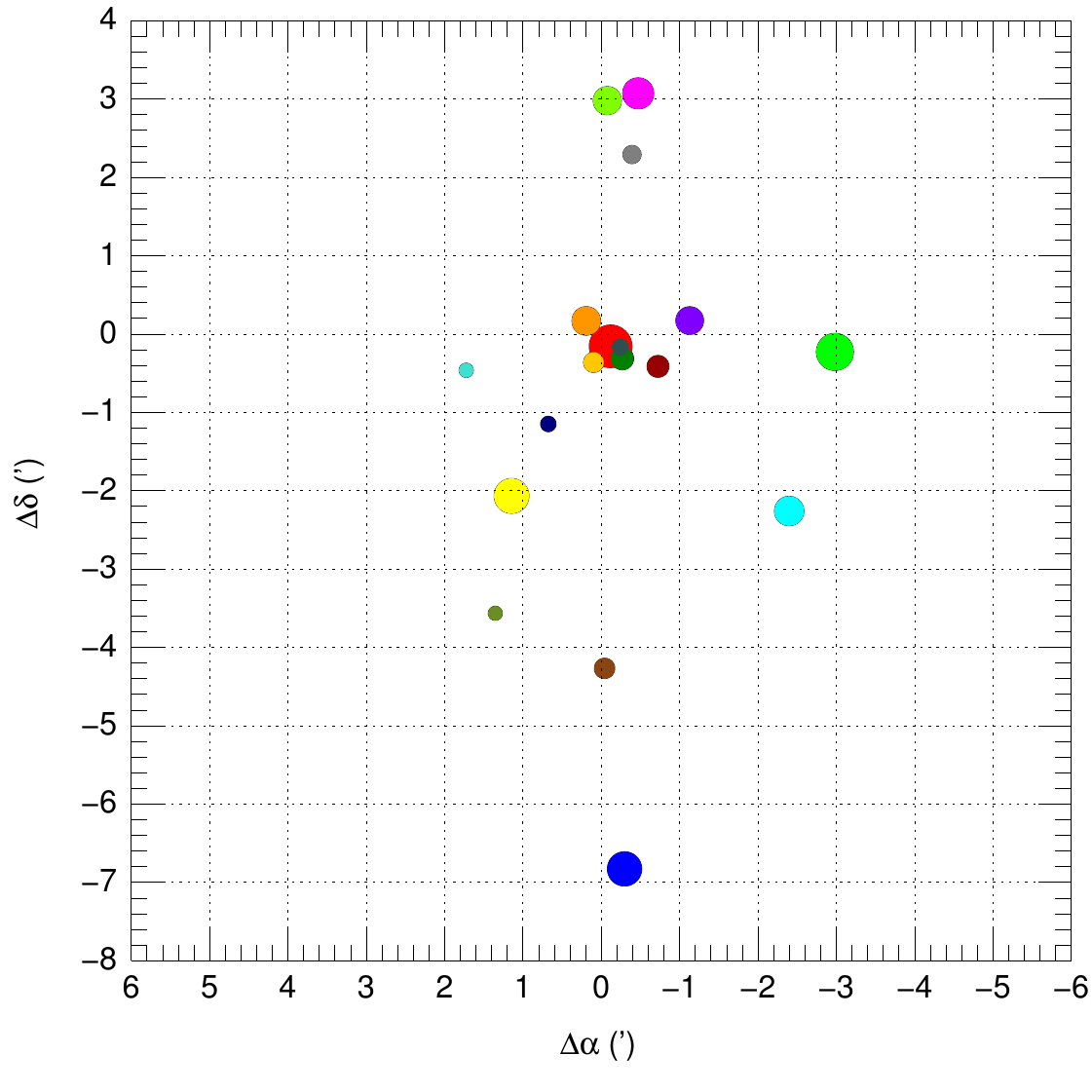}$\!$
             \includegraphics*[width=0.300\linewidth]{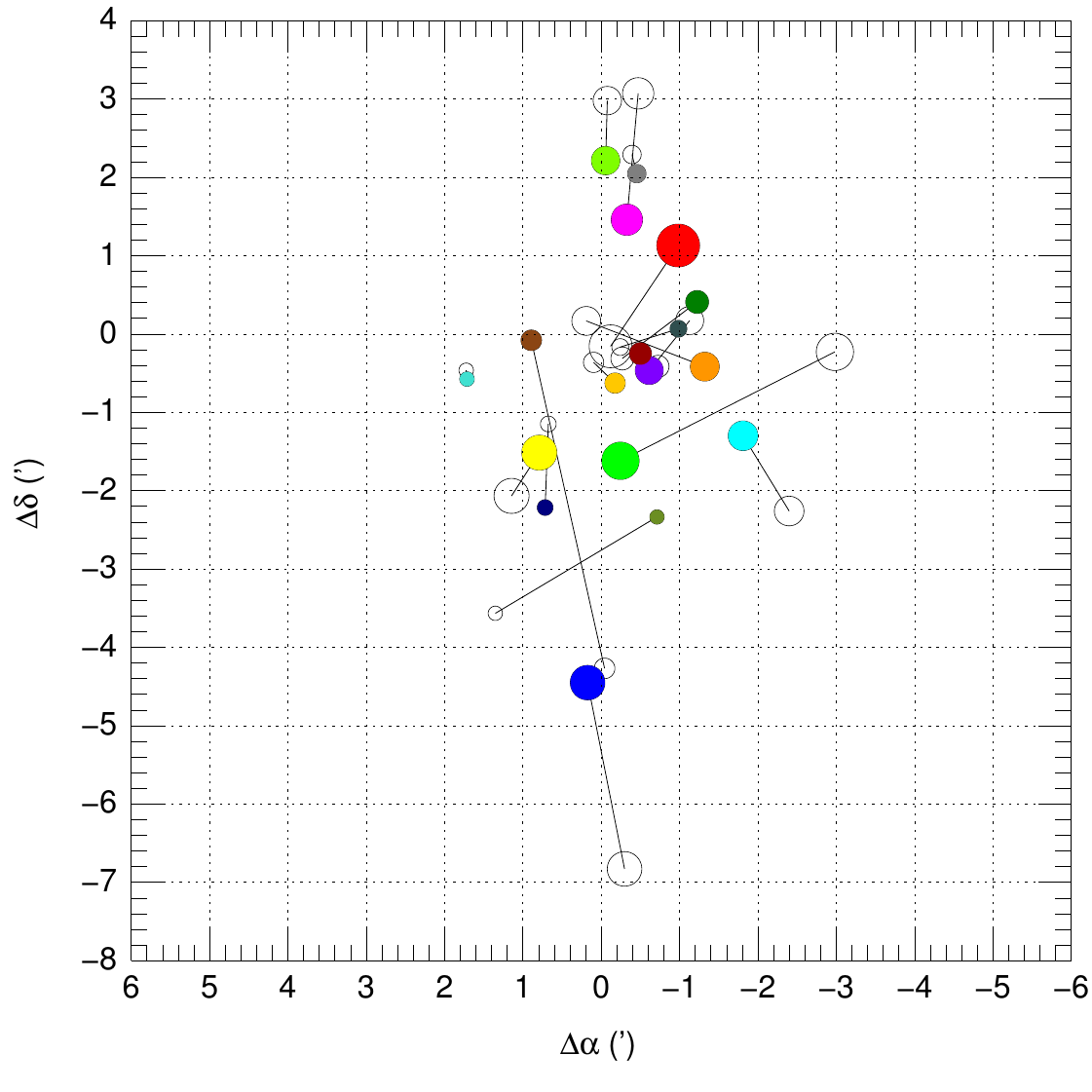}$\!$
             \includegraphics*[width=0.300\linewidth]{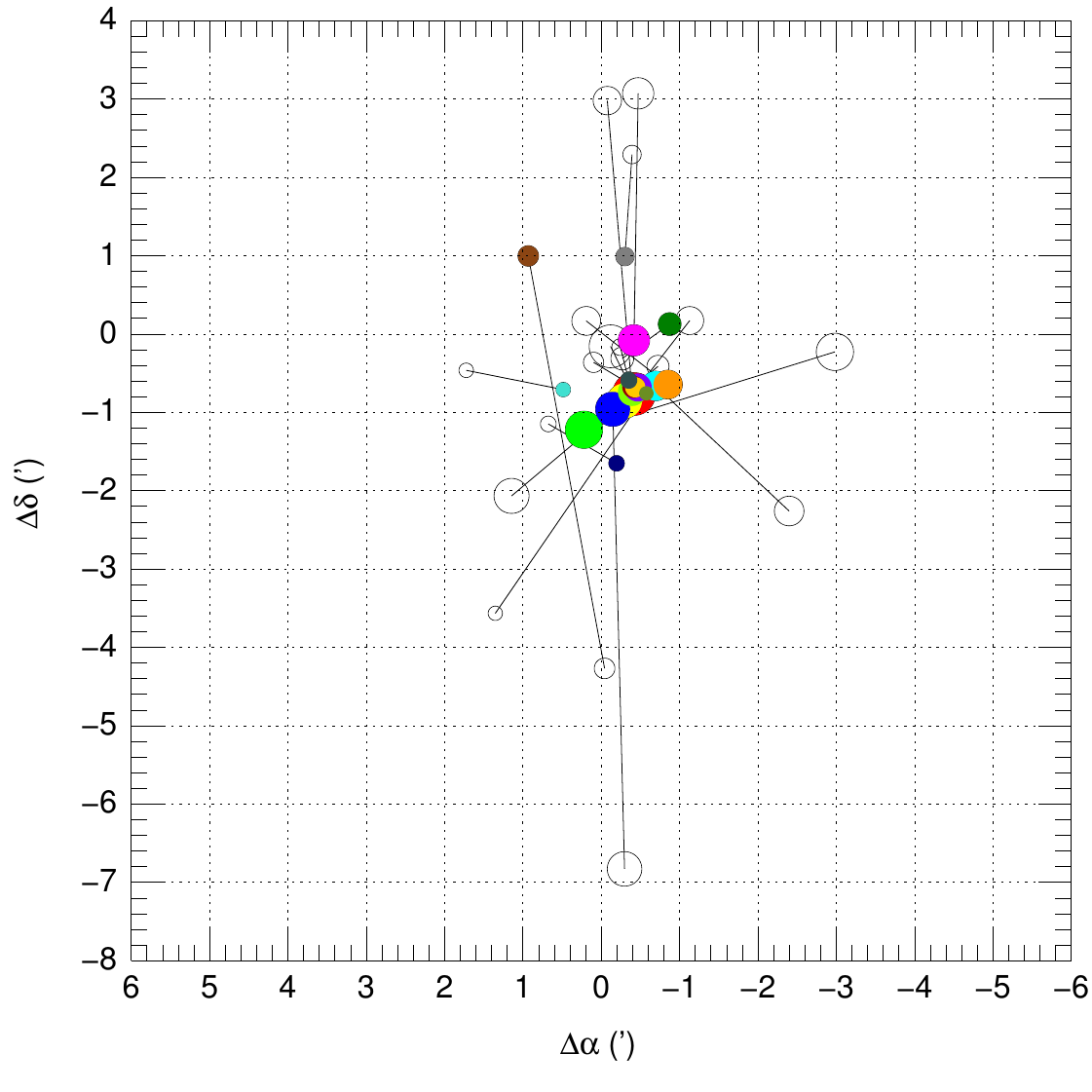}}
\caption{Charts for the 18 bright stars described in the text color-coded by ID and with symbol size representing a \GGGc\ magnitude scale. (left) Current positions. (center) Backtraced positions using the \textit{Gaia}~DR3 proper motions for 0.61~Ma ago, with empty circles used for the current positions and lines joining the backtraced and current positions. (right) Same as the previous panel for 0.88~Ma ago selecting the proper motions within 2.5~sigmas of the measured values that minimize the average mean separation between pairs.}
 \label{backtrack}   
\end{figure*}

$\,\!$\indent In order to estimate how the Galactic population at a distance similar to Stock~18 contaminates samples~1~to~3, we proceed along the following steps:

\begin{enumerate}
 \item We assume that the Galactic population has a proper-motion distribution that can be described by an inclined 2-D Gaussian distribution. We fit its parameters using the left panel of Fig~\ref{Gaia_pm} eliminating the white-circle region from the fit.
 \item Once we have a fit, we calculate how many stars from the Galactic population are in the region inside the circle to estimate the contamination for sample~3 and then we scale those numbers proportionally to the area for samples~1~and~2 assuming a spatially uniform distribution of the contamination.
 \item To estimate the probability of a star in sample~1 (or 2~or~3) of being a cluster member, we assume that the contaminants in the top panels of Fig.~\ref{Gaia_CAMDs} have the same distribution in the CAMD as in the lower left panel. Each star is then assigned a probability $p$ depending on the contaminant CAMD density in such a way that the total probability equals the expected number cleaned of contaminants. 
\end{enumerate}

The result of the second step is that $95\pm12$ stars in sample~1 are cluster members. There is significant contamination but $\sim$60\% of the stars remain in the cleaned sample. The results for the differential 1-2 and 2-3 samples are very different. In the first case, almost no stars remain ($6\pm17$) and in the second one only $26\pm21$~stars do: only 10\% of the stars in the differential 1-3 sample are cluster members, so the vast majority of the objects beyond 4\arcmin\ are just part of the Galactic population. A hint of this is seen in the right panel of Fig.~\ref{Gaia_CAMDs}: stars in sample 1 show a significant concentration in proper motion but stars in differential sample 1-3 show a smooth distribution. Therefore, \textit{Stock~18 is a compact cluster with a small or non-existent halo around it}, an issue that is further explored below.

The result of the third step for sample~1 is shown in the bottom right panel of Fig.~\ref{Gaia_CAMDs}. For the final (F) sample that we employ for the IMF calculation below, we use sample~1 with four additions: (a) GLS~\num{13370}~B (excluded from sample 1 due to its lack of \BPRP), for which we use the existing \GGGc\ value and estimate \BPRP\ assuming the same extinction as for A; (b) GLS~\num{13370}~Ab (excluded from sample 1 due to its non-detection in \textit{Gaia}~DR3), for which we estimate \GGGc\ from $\Delta z$ and \BPRP\ from assuming the same extinction as for A; and (c) GLS~\num{13368} and 2MASS~J00013558+6433174 (= Gaia~EDR3~\num{431768599809746688}), the two brightest blue stars in the differential 1-2 sample. They are likely to be cluster members for reasons that are explained in the next section.

\section{Internal cluster dynamics and structure}

\subsection{Dynamical state}

$\,\!$\indent Now we that have a good idea of which objects are likely cluster members, we analyze the cluster dynamics and structure. In Fig.~\ref{Gaia_chart_4} we plot the positions of the stars in sample~1, using the probability color coding developed before and with the relative proper motions plotted for the 16 brightest stars. Proper motions are not plotted for fainter stars because, in general, they have larger uncertainties. In that respect, we substitute the proper motion of the high-RUWE GLS~\num{13370}~A by that of GLS~\num{13370}~B, as both stars are expected to form a bound system with an orbit long enough for the two proper motions to be indistinguishable within the \textit{Gaia} time baseline.

\begin{table*}
\caption{Monte Carlo results for the proper motions and predicted positions 0.88~Ma ago for the 18~star sample. Measured and predicted proper motions are relative to the mean value of sample~1 (\pmrag, \pmdecg) and the predicted positions are relative to the cluster center (\alphac, \deltac). The measured proper motions of GLS~\num{13370}~A are those of GLS~\num{13370}~B.}
\label{backtrack_table}
\centerline{
\begin{tabular}{llrrrrrr}
\hline
\mci{Name} & \mci{\textit{Gaia}~DR3~ID} & \mci{\pmra} & \mci{\pmdec} & \mci{\pmra} & \mci{\pmdec} & \mci{$\Delta\alpha$} & \mci{$\Delta\delta$} \\
           &                            & (mas/a)     & (mas/a)      & (mas/a)     & (mas/a)      & \mci{(\arcmin)}      & \mci{(\arcmin)}      \\
           &                            & \mcii{measured}            & \mcii{predicted}           & \mcii{predicted}                            \\
\hline
GLS \num{13370} A         & \num{431769183925254272} &  0.085 & -0.127 &  0.020 &  0.042 & -0.419 & -0.766 \\
GLS \num{13362}           & \num{431771863984836352} & -0.269 &  0.137 & -0.218 &  0.068 &  0.219 & -1.223 \\
GLS \num{13376}           & \num{431768943407111424} &  0.035 & -0.054 &  0.098 & -0.083 & -0.297 & -0.854 \\
GLS \num{13368}           & \num{431766950542328448} & -0.046 & -0.234 & -0.010 & -0.400 & -0.149 & -0.960 \\
GLS \num{22355}           & \num{431775334318395904} & -0.014 &  0.159 & -0.004 &  0.215 & -0.419 & -0.080 \\
GLS \num{13365}           & \num{431771623466692608} & -0.058 & -0.094 & -0.116 & -0.109 & -0.696 & -0.663 \\
GLS \num{22356}           & \num{431769287004467584} &  0.149 &  0.057 &  0.071 &  0.055 & -0.853 & -0.645 \\
2MASS J00013526$+$6440324 & \num{431775330015871872} & -0.002 &  0.075 &  0.022 &  0.253 & -0.395 & -0.738 \\
2MASS J00012544$+$6437439 & \num{431769252644723456} & -0.051 &  0.063 & -0.045 &  0.058 & -0.467 & -0.683 \\
GLS \num{13370} D         & \num{431769183925255680} &  0.094 & -0.071 &  0.041 & -0.030 & -0.875 &  0.127 \\
2MASS J00012922$+$6437088 & \num{431769149565516288} & -0.022 & -0.016 & -0.021 &  0.018 & -0.411 & -0.676 \\
2MASS J00013558$+$6433174 & \num{431768599809746688} & -0.092 & -0.412 & -0.066 & -0.394 &  0.930 &  0.996 \\
GLS \num{13370} E         & \num{431769183925258496} &  0.027 &  0.026 &  0.036 &  0.022 & -0.435 & -0.680 \\
2MASS J00013231$+$6439510 & \num{431775231239186048} &  0.006 &  0.024 & -0.006 &  0.089 & -0.304 &  0.988 \\
GLS \num{13370} C         & \num{431769179623166976} &  0.073 & -0.023 &  0.007 &  0.029 & -0.353 & -0.591 \\
2MASS J00014229$+$6436247 & \num{431768977766839168} & -0.004 &  0.105 &  0.059 &  0.034 & -0.197 & -1.651 \\
2MASS J00015206$+$6437056 & \num{431769046486312320} &  0.001 &  0.011 &  0.084 &  0.017 &  0.484 & -0.710 \\
2MASS J00014859$+$6433596 & \num{431768458064599040} &  0.203 & -0.121 &  0.131 & -0.191 & -0.577 & -0.757 \\
\hline
\end{tabular}
}
\end{table*}

Two observations can be made about Fig.~\ref{Gaia_chart_4}. First is the difference between the spatial distribution of high and low probability members. The core is clearly dominated by high-probability members while the low-probability members are dispersed throughout the field. This probability-based difference validates the decontamination procedure of the previous section. Second, the plotted relative proper motions are not randomly distributed but show a tendency towards an outward radial direction, which is an indication of an expanding motion. Furthermore, the two brightest stars in the differential sample 1-2 (see the left panel of Fig.~\ref{Gaia_charts}, they are located in the lower part of the region, see above regarding their inclusion in the final sample) also form part of the same pattern. In order to see the effect better, in Fig.~\ref{Gaia_pm2} we plot coordinate vs. proper motion for 18 stars, the 16 brightest from sample~1 plus those two brightest from differential sample~1-2, which  corresponds to objects with $\gtrsim$~2~\Msun. An expansion pattern is clearly seen and the inverse of the slopes of the fitted linear regressions yield dynamical (interaction-free expansion) ages of $1.06\pm 0.11$~Ma in right ascension and $1.04\pm0.06$~Ma in declination, using as uncertainties the external proper motions without the systematic uncertainty (as we are only interested in the relative motions and the angular dispersion in the sky is small compared to the small-angle wavelength of $\sim 1^\circ$ seen in the \textit{Gaia}~DR3 angular covariance, \citealt{Maizetal21c}). The reduced $\chi^2$ values for the two fits are large (7.7 and 11.4 respectively), indicating that the expansion cannot be fully explained by a single event with no further interactions. However, the two plots in Fig.~\ref{Gaia_pm2} treat each coordinate independently. This is just an indication of an expansion but we need to confirm it with a 2-D analysis (as we cannot do a 3-D analysis for lack of accurate relative distances and radial velocities), which is what we do next. 

The left panel of Fig.~\ref{backtrack} shows the current positions of the 18~star~sample. The mean (plane-of-the-sky) separation of the 153 stellar pairs is 3.28\arcmin\ (2.77~pc) and the mean distance of the 18 stars to their centroid is 2.16\arcmin\ (1.83~pc). We use the relative proper motions to backtrace the positions in time (assuming no gravitational interactions) and we find the time at which the mean separation is a minimum. The result is 0.61~Ma ago, when the mean pair separation was 2.35\arcmin\ (1.99~pc) and the mean distance of the 18 stars to their centroid was 1.58\arcmin\ (1.34~pc), with the result plotted in the center panel of Fig.~\ref{backtrack}. At that point the 18 stars clearly form a more compact configuration, as in general that panel shows empty circles (current positions) surrounding a mixture of empty and colored circles. However, the core itself appears to be not as dense as it is at the current time. This is likely a combination of two effects: the gravitational interactions at the core curving the real trajectories into orbits and the consideration that the proper motions are exact, while the reality is that they have significant external uncertainties in \textit{Gaia}~DR3.

To address the uncertainty effect above we build a gradient descent Monte Carlo method to test how close the stars could have been in the past by varying the proper motions of the 18~stars. To include to some degree the effect of curvature, we
allow the proper motion in each coordinate to change within 2.5~sigmas of its average (using the external uncertainties without systematic uncertainties, see above), and we search for the point in time where the mean pair separation is minimized.
The result, shown in the right panel of Fig.~\ref{backtrack} is that 0.88~Ma ago the mean separation could have been as low as 0.91\arcmin\ (0.77~pc), with a corresponding mean distance of the 18 stars to their centroid of 0.61\arcmin\ (0.52~pc).
Furthermore, even though some of the 18~objects (e.g. 2MASS~J00013558+6433174) have trajectories that only graze the compact core at a distance of a few tenths of a pc, all of the massive stars in the system have trajectories that bring them
within 0.1~pc of the core either at that age or in the range 0.8-1.1~Ma. Therefore, \textit{the data are compatible with all of the massive stars in Stock~18 being formed in an extremely compact core and being ejected within a short
($\sim$0.3~Ma) period of time around 1~Ma ago.} Currently only the GLS~\num{13370}~Aa,Ab,B system (represented in the 18~star sample as a single object) and GLS~\num{22356} remain in the central 1-pc region but originally the other confirmed five
massive stars could have been there as well and forming a significantly more compact system. This hypothesis may be tested with future \textit{Gaia} data releases (such as DR4), as the proper motion uncertainties are expected to improve. For that
purpose, we provide in Table~\ref{backtrack_table} a list of the currently measured proper motions and of the ones resulting from the Monte Carlo analysis, as well as the predicted positions from 0.88~Ma ago\footnote{Such a prediction is subject
to a global displacement due to the uncertainty in the subtracted cluster proper motion.}. If Stock~18 was indeed a very compact cluster $\sim$1~Ma ago, future \textit{Gaia} data releases should move the proper motions from the former to something close to the latter. Given their likely origin at or near the core of Stock~18 and as already anticipated above, for the final sample used to determine the IMF below we include the two stars from the differential 1-2 sample, GLS~\num{13368} and 2MASS~J00013558+6433174, the first a massive star with a trajectory compatible with forming part of the $\sim$1.0~Ma event and the second an intermediate-mass object with only a grazing trajectory with respect to the core but with a possible interaction with GLS~\num{13376} at a slightly later epoch.

\subsection{Cluster structure}

\begin{figure}
 \centerline{\includegraphics*[width=\linewidth]{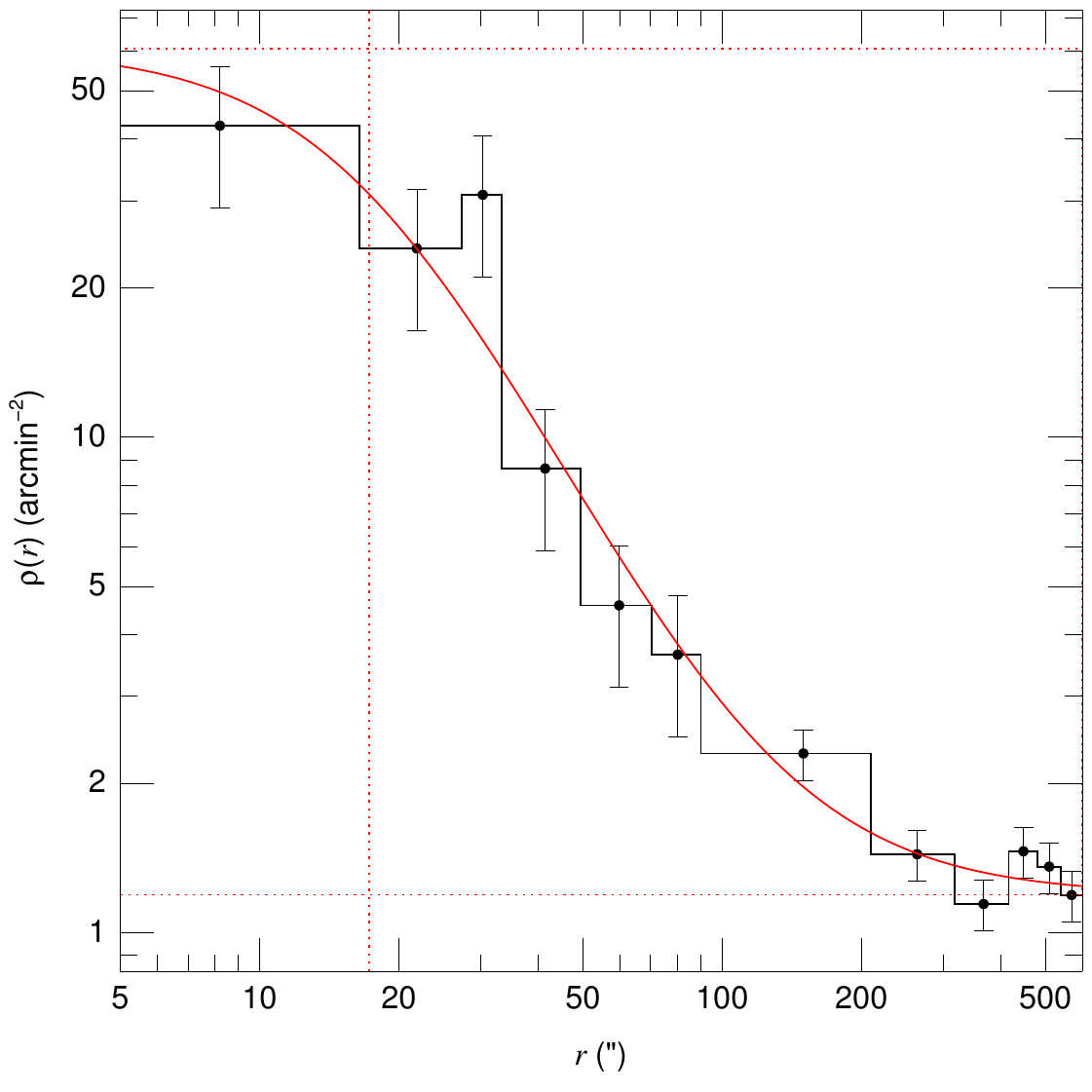}}
 \caption{Radial density profile for sample~3 (black) and fitted King profile (red). The dotted lines mark the fitted \rc, \fb, and \fb+\fc.}
 \label{king}   
\end{figure}

$\,\!$\indent The analysis in the previous subsection was centered on the brightest stars in the cluster but, of course, the cluster is dominated in numbers by low-mass stars. To study the overall cluster structure, we fit a King profile:

\begin {equation}
\rho(r)= \fb + \frac{\fc}{1+ (r/\rc)^2}\,
\label{King}
\end {equation}

\noindent where $\rho(r)$ is the stellar number density, \fb\ and \fc\ the background and cluster central number densities, and \rc\ the cluster core radius, to sample~3 as defined above. We do so by creating a histogram in the radial distribution, integrating Eqn.~\ref{King} in each bin, and fitting the result using a custom-made IDL procedure. It is known that dividing data into bins can create biases in the fitted function parameters and to avoid them we adopted three complementary strategies. [a] We divided the bins in two groups, one for the core region and one for the background and in each of those groups we select bins with the same number of stars in each, as this is known to significantly reduce biases \citep{MaizUbed05}. [b] After obtaining a first result, a second (final) iteration was done using as weights the values derived from the fitted function (as opposed to those derived from the data). [c] Different numbers of bins and different limits between the core and background regions were used to test for variations and to check for the validity of the uncertainties. The procedure was repeated in a 2-D grid in RA+dec for the center coordinates, with the final value selected by minimizing the value of \rc.

Using sample~3, we find that the cluster center is located at $\alphac = 0.400^{\circ}$, $\deltac = 64.626^{\circ}$, in each case with an uncertainty of $0.001^{\circ}$, and those are the values we use throughout this paper (Table~\ref{samples}). As for the parameters of the King distribution, we obtain $\rc = 0.29\pm 0.08\arcmin$, $\fc = 60\pm 24$~arcmin$^{-2}$, and $\fc = 1.2\pm 0.1$~arcmin$^{-2}$. Our value for the cluster core is similar to the 2MASS value derived by \citet{bhatt2012stellar} but lower than the rest in Table~\ref{Stock18_results}.

To check for possible issues with the sample selection, we repeated the procedure using different samples. First, we used the same sample~3 but with different cuts in \GGG\ to select only low-mass stars. Second, we substitute sample~3 by the objects in the final sample with $p > 0.5$ (and fixing \fb\ to a value of zero). In all cases we find similar values for the core radius, with variations much smaller than the uncertainty quoted above.

\subsection{Dynamical analysis}

$\,\!$\indent The core radius we determine for Stock~18 of $0.24\pm 0.07$~pc would be one of the smallest in the large sample of Galactic clusters of \citet{kharchenko2013global}. The compact size would be remarkable by itself but it is even more so considering our discovery of the expansion of the high+intermediate mass stars from the previous subsection, which would indicate an even more compact size at birth. The outstanding question is whether the expansion detected also affects the low-mass stars. This cannot be tested directly using their proper motions at the present time, as their uncertainties in \textit{Gaia}~DR3 are, in general, too large. We can test it indirectly by determining the mean separation and the mean distance to their centroid for the final sample stars with $p > 0.8$, which should be a relatively clean sample of cluster members dominated by low-mass stars. Those values are 2.16\arcmin\ and 1.66\arcmin, respectively, that is, significantly smaller that the ones for the current positions of the 18~star sample and relatively similar to the ones of that same sample 0.61~Ma ago assuming exact proper motions. Therefore, apparently \textit{the expansion of the more massive stars in Stock~18 is not shared by the stars of lower mass.}

What mechanism could be responsible for such behavior? In the classical view of \citet{LadaLada03}, most clusters go through an embedded, compact phase, after which they lose most of their associated gas and become unbound as a result. However, this does not appear to be the case for two reasons: first, because the gas appears to be cospatial with the stars and, second, because a global mass loss should affect both high- and low-mass stars equally, which does not appear to be the case for Stock~18. Indeed, a recent analysis by \citet{Mireetal24} has found an average delay of 5.5~Ma between cluster formation and the onset of expansion, so it is likely too early for that to have happened. Given the very young age of Stock~18 (see below) and compactness, other mechanisms such as tidal disruption cannot be at work, either.

One process that preferentially affects massive stars are $n$-body interactions with $n\ge 3$ \citep{Ohetal15,OhKrou16}, since when those take place among low-mass stars alone the resulting outgoing velocities are unlikely to be above the escape velocity of the cluster. In our recent analysis of the Bermuda cluster (\VO{}{014}~NW, \citealt{Maizetal22b}) we discovered how multiple stellar encounters between massive stars were able to eject so much mass in the form of high-mass systems as to provoke the expansion of the remaining low-mass stars, thus leaving and expanding cluster orphaned of massive stars. The most likely explanation for Stock~18 consistent with the data in this paper is that \textit{one or several events $\sim$~1~Ma ago ejected some of the massive stars in the cluster but without enough mass being lost as to unbind the cluster.} Therefore, this is the second cluster for which we find that dynamical interactions between massive stars have radically altered the dynamics of the system at a very early age, before gas loss is unable to contribute in a significant manner. As we pointed out in \citet{Maizetal22b}, there are hints of other clusters where this may have happened as well, such as in \VO{012}~S (Haffner~18) and \VO{023} (Orion Nebula Cluster). One difference between Stock~18 and the other three is its lower mass, both for the cluster itself and for the most massive star. That is the likely reason for the ejected stars to have a likely low ejection velocity, thus producing only walkway stars but not runaways\footnote{We say likely because we are not including radial velocities in our analysis.}. 

\section{Age and variability}

$\,\!$\indent As we mentioned in the introduction, previous age estimates indicate that Stock~18 is a very young cluster. \citet{roger2004sharpless} analyzed the \HII\ region to give an age of 0.25-0.50 since the activation of the main ionizing source (GLS~\num{13379}~Aa) and \citet{sinha2020variable} detected a peak in the PMS distribution at an age of 1~Ma, with some older stars. In the previous section we have obtained a dynamical age of 1~Ma with an uncertainty of just 0.1-0.2~Ma. The good agreement with the \citet{sinha2020variable} points towards Stock~18 being created in such a compact state that a dynamical instability took place almost immediately. If all the massive stars were indeed inside a region as small as 0.1~pc as suggested in the previous section, the numerical simulations of \citet{OhKrou16} suggest that a large fraction of the massive stars may be ejected.

We can provide an independent estimate of the PMS age by analyzing the lower right panel of Fig~\ref{Gaia_CAMDs}, where we have plotted the 1.0~and~2.0~Ma isochrones of \citet{Baraetal15}. As it turns out, for stars around $\sim$1~\Msun\ differential extinction (unknown a priori for individual objects) shifts the isochrones in the CAMD in a direction nearly parallel to the isochrones themselves. This makes age relatively easy to determine and mass more difficult. For the stars with $p > 0.8$ we confirm the finding of \citet{sinha2020variable} of a peak around an age of $\sim$1~Ma and we also find a handful of stars significantly younger than those. Stars with probabilities between 0.5 and 0.8 have somewhat older ages, as expected from the decontamination procedure. Therefore, our analysis confirms the young age of the PMS population.

Figure~\ref{Gaia_CAMDs} and Table~\ref{Gaia_stars} can also be used to analyze the variability of the samples based on the values of \sG\ from \citet{Maizetal23}. Stars in Table~\ref{Gaia_stars} have low variability, as expected from stars in the upper MS that are not eclipsing binaries or of Oe/Be type. There is, however, a correlation with luminosity as we go from mid-B to O stars in the MS, as seen in the large sample of \citet{Maizetal23}.

The variability analysis of Fig.~\ref{Gaia_CAMDs} is limited to $\GGG < 17$ by the input data in \citet{Maizetal23}. Most stars with \sG\ values have low variability. The more variable stars in sample~1, with the exception of GLS~\num{13370}~A (see above), are all intermediate-mass stars of $\sim$~2~\Msun\ at an intermediate point between the PMS and MS isochrones, that is, they are on their final stages before reaching the main sequence. As pointed out in \citet{Maizetal23}, the high \sG\ values can be used as an approximate proxy to distinguish stars that have not yet reached the MS. Comparing that region of the CAMD with the equivalent ones for samples 1-3 and 3-4, we see that in the latter two most stars do not have high variability but a few do. Those may be the few PMS in the halo of Stock~18. In addition, those two samples also show stars further to the right (about $\sim$~1 mag in \BPRP) crossing in diagonal from the top left to the lower right. Those are likely red clump stars of different extinctions at a distance similar to that of Stock~18 and are also characterized by low values of \sG\ \citep{Maizetal23}. Finally, one star with high variability appears towards the upper right corner of the lower left panel of Fig.~\ref{Gaia_CAMDs}, 2MASS~J00003565$+$6440364 and is likely a red giant. 

\section{IMF and cluster mass}

\begin{table}
\caption{Number of stars per mass range and expected values from a canonical IMF assuming the same number of stars in the intermediate mass range.}
\label{masses}
\centerline{
\begin{tabular}{rrrrrr}
\hline
\mci{mass}    & \mci{$n$}     & \mci{$n$}     & \mci{$n$} & \mci{$m$}     & \mci{$\overline{m}$} \\
\mci{range}   & \mci{$p>0.8$} & \mci{$p>0.5$} & \mciii{------ canonical -------}                 \\
\mci{(\Msun)} &               &               &           & \mci{(\Msun)} & \mci{(\Msun)}        \\
\hline
  $>$8.0      &             9 &             9 &       2.7 &          52.6 &                19.13 \\
 1.8-8.0      &            17 &            17 &      17.0 &          55.9 &                 3.29 \\
 0.7-1.8      &            42 &            81 &      47.9 &          50.7 &                 1.06 \\
0.08-0.7      &           --- &           --- &     370.6 &          92.3 &                 0.25 \\
Total         &           --- &           --- &     438.2 &         251.5 &                 0.57 \\
\hline
\end{tabular}
}
\end{table}

$\,\!$\indent The next step in our analysis of Stock~18 is the calculation of its IMF and total stellar mass. If extinction were (approximately) constant across the field, we could simply use the CAMD to determine the individual stellar masses using either the MS or PMS isochrones. However, as we have previously seen, there is significant differential extinction, which introduces a corresponding uncertainty in the stellar masses. This can be seen in Fig.~\ref{Gaia_CAMDs}, where pairs of isochrones spanning the range of extinction are shown and some initial masses are plotted in the MS isochrones for reference. As a further complication, due to the young age of the cluster, intermediate-mass stars (defined as the 1.5-8.0~\Msun\ range) are in between the PMS and MS isochrones. The only case for which we can derive individual realistic stellar masses are for the objects for which we have spectral classifications.

Given that, we performed a simplified analysis by dividing our sample into three observed bins: massive stars ($M > 8.0$~\Msun), intermediate-mass stars ($M = 1.5-8.0$~\Msun), and low-mass stars ($M = 0.7-1.5$~\Msun), with a fourth bin of very low mass stars ($M = 0.08-0.7$~\Msun) not observed. We then counted how many stars are in each bin and compared the results with what would be expected from a canonical IMF \citep{Krou01}, see Table~\ref{masses}. An analysis of that table reveals:

\begin{itemize}
 \item All stars in the first two bins have $p > 0.8$.
 \item If the IMF follows the canonical one for intermediate- and low-mass stars, then most of the stars with $p < 0.8$ are not cluster members. Slightly changing the 0.7~\Msun\ value for the lowest mass observed or introducing a reasonable completeness function does not change this assessment. 
 \item The observed number of high-mass stars is incompatible with the expected number from a canonical IMF with 17 stars, indicating that \textit{the IMF is top-heavy}.  
 \item If we consider the ejected stars as no longer part of the system, then the PDMF is compatible with a canonical one.
 \item If we estimate the masses of the stars in the high-mass range from the CHORIZOS analysis we get a total mass that is higher by a factor of two ($101\pm 4$~\Msun) but a significantly lower average mass (11.3~\Msun) in comparison with the values of the canonical IMF. In other words, the IMF is top heavy but only for massive stars that are right above the 8~\Msun\ limit. 
 \item Adopting a canonical IMF for stars with $M < 8$~\Msun\ and the individual masses for $M > 8$~\Msun\ we arrive to a total cluster mass of $\sim$300~\Msun\ in the 0.08-25~\Msun\ range (no stars with higher masses are present), of which 1/3 is in massive stars.
\end{itemize}

It is interesting to compare these results with the previous ones we obtained for the Bermuda cluster in \citet{Maizetal22b}. Both are relatively low-mass clusters with a top-heavy IMF that have ejected a significant fraction of their massive stars, thus significantly reducing the cluster mass. There are differences in the IMF details (the Bermuda cluster has produced stars with higher masses, most noticeably the primary in the Bajamar system), the ejection speed (the Bermuda cluster has produced true runaways as opposed to just walkaways) and the age (the Bermuda cluster is about twice as old) but two consequences remain the same: \textit{a top-heavy IMF turns into a nearly canonical PDMF prior to any supernova explosions due to dynamical interactions and the number of free-floating compact objects in the Galaxy increases significantly.} Both of those issues have important consequences for Galactic modelling, so further studies need to be pursued to establish how frequent is this phenomenon, that may have been overlooked in the past due to data of sufficient quality not being available. 

\section{Galactic location}

$\,\!$\indent The last aspect we analyze in this paper is the motion of Stock~18 with respect to the Milky Way. In Table~\ref{sample_results} the proper motions in RA and declination of sample~1 are $-2.670\pm0.024$~mas/a and $-0.658\pm0.024$~mas/a, respectively. From our previous analysis of the Galactic population at the distance of Stock~18 (sample~4 discarding the proper motion region around Stock~18 and fitting a 2-D Gaussian distribution) we obtain values of $-2.008\pm0.030$~mas/a and $-0.552\pm0.019$~mas/a. Therefore, Stock~18 has a distinct group motion with respect to the surrounding population, as the differences are many times larger than the uncertainties. Translating those values into Galactic coordinates and assuming no covariance, we obtain values of $\pmlong = -2.747\pm0.024$~mas/a and $\pmlatg = -0.133\pm0.024$~mas/a for the cluster. The equivalent values for the Galactic population are $-2.077\pm0.030$~mas/a and $-0.156\pm0.020$~mas/a, respectively. At a distance of 2.91~kpc, those translate in Galactic coordinates into tangential velocities of ($-$37.9,$-$1.8) km/s and ($-$28.6,$-$2.1) km/s. Hence the cluster has a significant velocity of 9.2~km/s in the plane of the sky with respect to the surrounding population, with the vector pointing nearly parallel to the Galactic plane. 

How do the group proper motions of Stock~18 and the surrounding population compare to the expected one at its Galactic location? To calculate that, we use:

\begin{itemize}
 \item The height of the Sun above the Galactic Plane of \citet{Maizetal08a}, $z_\odot$~=~20~pc.
 \item The peculiar solar velocity with respect to the LSR of \citet{Schoetal10b}: $U_\odot$~=~11.1~km/s, $V_\odot$~=~12.24 km/s, and $W_\odot$~=~7.25~km/s.
 \item A distance to the Galactic Center of 8.178~kpc from \citet{Abutetal19}.
 \item The Galactic rotation of \citet{Sofu20} updated in 2021\footnote{\url{https://www.ioa.s.u-tokyo.ac.jp/~sofue/htdocs/2017paReview/MW-2017pasjReview.dat}}.
\end{itemize}

With those, we obtain $-2.449$~mas/a in $l$ and $-0.386$~mas/a in $b$ as the expected proper motions. At face value, with respect to the expected values in Galactic coordinates [a] Stock~18 is moving westward and the Galactic population is moving eastward and [b] both are moving northward. The difference between the Galactic population and the model in the longitude component is just 5.1~km/s, small enough that we cannot discard that the fault lies in the rotation curve model. On the other hand, the discrepancy between the cluster and the Galactic population is significant. One possible cause is the effect of the Perseus arm, where Stock~18 is located, on the motion of its natal cloud.  

We also point out that at the Galactic latitude of Stock~18 of 2.27$^\circ$, 2.91~kpc corresponds to a vertical distance of 115~pc, which added to the value of $z_\odot$ yields 135~pc above the Galactic mid-plane at our location. Therefore, the cluster is located above our Galactic mid-plane and moving away from it. This is likely an effect of the Galactic warp, which in this region of the Galaxy curves upwards \citep{Romeetal19}. Indeed, the three other clusters with O stars analyzed in Villafranca~I~and~II that are located at distances between 2~and~3 kpc and in the $l=84-135^\circ$ region (Berkeley~90, Sh~2-158, and IC~1805) are 54-201~pc above our mid-plane, so Stock~18 is not an exception.

\section{Summary and future work}

\begin{table*}
\caption{Summary of results in this paper. The fourth column refers to the section or table in the paper where the value is presented or derived.}
\label{summary}
\centerline{
\begin{tabular}{lrccl}
\hline
Parameter      &       \mci{Value}                 & Units  & S/T  & Reference, method, notes                                         \\
\hline
name           &       \mci{Stock~18}              & ---    & S1   & \citet{macconnell12006homage}                                    \\
               &       \mci{Sh 2-170}              & ---    & S1   & \citet{Shar59}                                                   \\
               &       \mci{\VO{036}}              & ---    & S2.3 & ---                                                              \\
\alphac        &        0.400$\pm$0.001            & deg    & S7.2 & \rc\ minimization                                                \\
\deltac        &       64.626$\pm$0.001            & deg    & S7.2 & \rc\ minimization                                                \\
\hline
\pmrag         &       $-$2.670$\pm$0.024          & mas/a  & T8   & Villafranca method, sample 1                                     \\
\pmdecg        &       $-$0.658$\pm$0.024          & mas/a  & T8   & Villafranca method, sample 1                                     \\
\pmlong        &       $-$2.747$\pm$0.024          & mas/a  & S10  & Villafranca method, sample 1                                     \\
\pmlatg        &       $-$0.133$\pm$0.024          & mas/a  & S10  & Villafranca method, sample 1                                     \\
\pmrag         &       $-$2.008$\pm$0.030          & mas/a  & S6.3 & 2-D Gaussian fitting, Galactic population at same distance       \\
\pmdecg        &       $-$0.552$\pm$0.019          & mas/a  & S6.3 & 2-D Gaussian fitting, Galactic population at same distance       \\
\pmlong        &       $-$2.077$\pm$0.030          & mas/a  & S10  & 2-D Gaussian fitting, Galactic population at same distance       \\
\pmlatg        &       $-$0.156$\pm$0.020          & mas/a  & S10  & 2-D Gaussian fitting, Galactic population at same distance       \\
$d$            &       2.91$\pm$0.10\phantom{0}    & kpc    & T8   & Villafranca method, sample 1                                     \\
\hline
age            &       1.0\phantom{0$\pm$0.070}    & Ma     & S1   & \citet{sinha2020variable}, PMS population                        \\
               &       0.25$-$0.50\phantom{0}      & Ma     & S1   & \citet{roger2004sharpless}, \HII\ region                         \\
               &       0.80$-$1.10\phantom{0}      & Ma     & S7.1 & dynamical expansion of massive stars                             \\
               &       1.0\phantom{0$\pm$0.070}    & Ma     & S8   & PMS population                                                   \\
\hline
mass           & $\sim$300\phantom{.40$\pm$0.070}  & \Msun  & S9   & 0.08$-$25 \Msun\ range, individual masses plus IMF extrapolation \\
IMF            & \mci{top heavy}                   & ---    & S9   & $M > 8$~\Msun, ratio of high- to intermediate-mass stars         \\
\hline
\rc            &       0.28$\pm$0.08\phantom{0}    & arcmin & S7.2 & King profile fitting                                             \\
               &       0.24$\pm$0.07\phantom{0}    & pc     & S7.2 & King profile fitting + \textit{Gaia} DR3 distance                \\
\hline
\EBV           &       0.55$-$0.81\phantom{0}      & mag    & S5   & CHORIZOS                                                         \\
\RV            &       3.15$-$3.80\phantom{0}      & ---    & S5   & CHORIZOS                                                         \\
\AV            &       1.83$-$2.58\phantom{0}      & mag    & S5   & CHORIZOS                                                         \\
\AG            &       1.77$-$2.43\phantom{0}      & mag    & S5   & CHORIZOS                                                         \\
\hline
spectral clas. & O9\phantom{.5:} V\phantom{nnn}    & ---    & T3   & GLS~\num{13370}~A, MGB, dominant ionizing source                 \\
               & B1\phantom{.5:} V\phantom{nnn}    & ---    & T3   & GLS~\num{13362}, MGB                                             \\
               & B0.5\phantom{:} V(n)\phantom{(}   & ---    & T3   & GLS~\num{13376}, MGB                                             \\
               & B0.5\phantom{:} V\phantom{nnn}    & ---    & T3   & GLS~\num{13368}, MGB                                             \\
               & B1\phantom{.5:} V\phantom{nnn}    & ---    & T3   & GLS~\num{22355}, MGB                                             \\
               & B1.5\phantom{:} V\phantom{nnn}    & ---    & T3   & GLS~\num{13365}, MGB                                             \\
               & B1:\phantom{.5} V\phantom{nnn}    & ---    & T3   & GLS~\num{13370}~B, MGB                                           \\
               & B2\phantom{.5:} V\phantom{nnn}    & ---    & T3   & GLS~\num{22356}, MGB                                             \\
               & B2.5: Vnnn                        & ---    & T3   & 2MASS~J00012544$+$6437439, MGB                                   \\
\hline
separation     &         0.49\phantom{$\pm$0.070}  & arcsec & T2   & GLS~\num{13370}~Aa,Ab, AstraLux PSF fitting                      \\
PA             &       267.4\phantom{0$\pm$0.070}  & deg    & T2   & GLS~\num{13370}~Aa,Ab, AstraLux PSF fitting                      \\
$\Delta z$     &         3.91\phantom{$\pm$0.070}  & mag    & T2   & GLS~\num{13370}~Aa,Ab, AstraLux PSF fitting                      \\
\hline
\end{tabular}
}
\end{table*}

$\,\!$\indent A summary of the results in this paper is given in Table~\ref{summary}. In more detail:

\begin{itemize}
 \item Stock~18 is a young stellar cluster located at a distance of 2.91$\pm$0.10~kpc, which in its Galactic direction corresponds to the Perseus arm.
 \item It is quite compact, with a core radius of 0.24$\pm$0.07~pc and no significant halo, but it was even more compact in the past. Its massive stars are expanding and currently are more disperse than its low- and intermediate-mass stars. It is possible that 0.8-1.1~Ma ago most of the massive stars were contained at one point within a space of less than 0.1~pc.
 \item We propose that dynamical interactions at that time, right after cluster formation, ejected most of the massive stars. If so, this would be the second cluster in the Villafranca sample (after the Bermuda cluster) for which such interactions have produced an expansion, albeit in this case restricted to just the massive stars and with overall low-ejection velocities (i.e. walkaways instead of runaways). The process takes place before the cluster loses most of its natal gas.
 \item The cluster IMF is top heavy, with 9 massive stars and only 2.7 expected from the number of intermediate-mass stars. However, if we exclude the ejected massive stars, the PDMF looks like a canonical one (or Kroupa) prior to any SN explosions. We observed this same effect in the Bermuda cluster. If confirmed for more clusters, it would have two important consequences: the true IMF would be top-heavy with respect to the canonical one, as prior calculations would have excluded the ejected stars, and the number of free-floating compact objects would be higher than expected, as the ejected stars will eventually explode as SNe.
 \item The total stellar mass of Stock~18 is $\sim$300~\Msun, of which $\sim$~1/3 is in massive stars.
 \item There is significant differential extinction in the cluster, which hampers the derivation of some of its parameters. To a lesser degree, \RV\ is also variable and, in general, larger than the canonical 3.1 value.
 \item Stock~18 is above the Galactic mid-plane, likely as a result of the Galactic warp,  and is moving in the plane of the sky with respect to the surrounding population at a speed of 9.2~km/s, possibly as a reflection of the motion of its natal cloud with respect to the Perseus arm.
 \item The ionizing flux of the cluster is dominated by the more massive component of the GLS~\num{13370} system, for which we have detected for the first time a new visual component (Ab) in our AstraLux data. 
\end{itemize}

The most important future line of work regarding Stock~18 will be redoing the dynamical analysis with \textit{Gaia}~DR4 to verify the hypothesis that the cluster was in a very compact state $\sim$1.0~Ma ago. We have also applied to get
time to observe the core of the cluster with the WEAVE LIFU, as that would allow us to obtain spectroscopy of more stars and to derive the gas extinction from the hydrogen emission lines. We will also continue with the Villafranca project to
attempt the detection of clusters in a similar state. However, we think it is unlikely that we will be able to catch a system in the act of being in a compact $\sim$0.1~pc state for two reasons: extinction is expected to be significantly higher
at very early stages (so it would only be observable in the IR) and, if \citet{OhKrou16} are right, such a phase would be so unstable that it could not last long, so we would only detect systems in the aftermath.

\begin{acknowledgements}
J.~M.~A., A.~S., and M.~P.~G. acknowledge support from the Spanish Government Ministerio de Ciencia e Innovaci\'on and Agencia Estatal de Investigaci\'on (\num{10.13039}/\num{501100011033}) through grant PID2022-\num{136640}~NB-C22 and from the Consejo Superior de Investigaciones Cient\'ificas (CSIC) through grant 2022-AEP~005. 
The authors extend their appreciation to the Deanship of Scientific Research at Northern Border University, Arar, KSA for funding this research work through the project number ``NBU-FPEJ-2024-237-01''.
This work has made use of data from the European Space Agency (ESA) mission \href{https://www.cosmos.esa.int/gaia}{\it Gaia}, processed by the {\it Gaia} Data Processing and Analysis Consortium (\href{https://www.cosmos.esa.int/web/gaia/dpac/consortium}{DPAC}). Funding for the DPAC has been provided by national institutions, in particular the institutions participating in the {\it Gaia} Multilateral Agreement. The {\it Gaia} data is processed with the computer resources at Mare Nostrum and the technical support provided by BSC-CNS. 
This work includes data obtained with the OSIRIS spectrograph at the 10.4~m Gran Telescopio de Canarias (GTC) and with the AstraLux lucky imaging instrument at the 2.2~m Telescope at Calar Alto (CAHA).
\end{acknowledgements}




\bibliographystyle{aa} 
\bibliography{example} 

$\,\!$

\vfill

\eject

\begin{appendix}

\section{Glossary}

$\,\!$\indent We provide a list of acronyms and terms used in this paper. 

\begin{itemize}
 \item \alphac, \deltac: Filtering central coordinates for cluster membership selection (cluster center), see Villafranca papers.  
 \item ALS: Alma Luminous Star.
 \item \AV: Monochromatic extinction at 5495~\AA, see \citet{Maiz24}.
 \item \AG: Extinction integrated over the \GGG\ band.
 \item ALS: Alma Luminous Star catalog, \cite{Reed03,Pantetal21}.
 \item \Cstar: Photometric contamination parameter from Eqn.~6 of \citet{Rieletal21}.
 \item CAMD: Color-Absolute Magnitude Diagram.
 \item CCD: Charged-Coupled Device.
 \item CDS: Centre de Donn\'ees astronomiques de Strasbourg.
 \item CHORIZOS: a \textbf{CH}i-square c\textbf{O}de for paramete\textbf{R}ized model\textbf{I}ng and characteri\textbf{Z}ation of ph\textbf{O}tometry and \textbf{S}pectrophotometry, \citet{Maiz04c}.
 \item CMD: Color-Magnitude Diagram.
 \item CSIC: Consejo Superior de Investigaciones Cient{\'\i}ficas.
 \item $d$: Group distance.
 \item $\Delta(\BPRP)$: Filtering color displacement with respect to isochrone, see Villafranca papers. 
 \item $\Delta m$: Magnitude difference in a binary system.
 \item \EBV: Monochromatic color excess between 4405~\AA\ and 5495~\AA, see \citet{Maiz24}.
 \item ESA: European Space Agency.
 \item \fc: Central source density for the King profile.
 \item \fb: Background source density for the King profile.
 \item \GG: \textit{Gaia}~DR2 $G$-band magnitude.
 \item \GGG: \textit{Gaia}~EDR3 $G$-band magnitude.
 \item \GGc: \textit{Gaia}~DR2 corrected $G$-band magnitude (Weiler et al. in prep.).
 \item \GGGc: \textit{Gaia}~EDR3 corrected $G$-band magnitude (Weiler et al. in prep.).
 \item \GGBP: \textit{Gaia}~DR2 BP-band magnitude.
 \item \GGGBP: \textit{Gaia}~EDR3 BP-band magnitude.
 \item \GGRP: \textit{Gaia}~DR2 RP-band magnitude.
 \item \GGGRP: \textit{Gaia}~EDR3 RP-band magnitude.
 \item \BPRP: \textit{Gaia}~EDR3 BP$-$RP color.
 \item \textit{Gaia} DR2: Second \textit{Gaia} Data Release.
 \item \textit{Gaia} EDR3: Early third \textit{Gaia} Data Release.
 \item \textit{Gaia} DR3: Third \textit{Gaia} Data Release.
 \item \textit{Gaia} DR4: Fourth \textit{Gaia} Data Release.
 \item GLS: Galactic Luminous Star.
 \item GOSC: Galactic O-Star Catalog, \citet{Maizetal04b}.
 \item GOSSS: Galactic O-Star Spectroscopic Survey, \citet{Maizetal11}.
 \item GTC: Gran Telescopio de Canarias.
 \item HRD: Hertzsprung-Russell Diagram.
 \item IMF: Initial mass Function.
 \item $k$: Multiplicative constant for parallax/proper motion uncertainties, see \citet{Maiz22} and Eqn.~\ref{sigmae}. 
 \item LC: Luminosity class, either from spectral classification or as fitted by CHORIZOS.
 \item LiLiMaRlin: \textbf{Li}brary of \textbf{Li}braries of \textbf{Ma}ssive-Star High-\textbf{R}eso\textbf{l}ut\textbf{i}o\textbf{n} Spectra, \citet{Maizetal19a}.
 \item LLS: LMC Luminous Star.
 \item LMC: Large Magellanic Cloud.
 \item LSR: Local Standard of Rest.
\end{itemize}

\vfill

\eject

\begin{itemize}
 \item \pmrac, \pmdecc: Filtering central proper motion for cluster membership selection, see Villafranca papers.       
 \item \pmrag, \pmdecg: Group proper motion in equatorial coordinates, see Villafranca papers.
 \item \pmlong, \pmlatg: Group proper motion in Galactic coordinates, see the \textit{Gaia} \href{https://gea.esac.esa.int/archive/documentation/GDR2/Data_processing/chap_cu3ast/sec_cu3ast_intro/ssec_cu3ast_intro_tansforms.html}{documentation}\footnote{ \url{https://gea.esac.esa.int/archive/documentation/GDR2/Data_processing/chap_cu3ast/sec_cu3ast_intro/ssec_cu3ast_intro_tansforms.html}}.
 \item \MG: Absolute magnitude in the \GGG\ band.
 \item MGB: Marxist Ghost Buster, \citet{Maizetal12,Maizetal15b}.
 \item MS: Main Sequence.
 \item MW: Milky Way.
 \item $N_{*,0}$: Number of stars in group before normalized parallax filtering, see Villafranca papers.
 \item $N_*$: Number of stars in group after normalized parallax filtering, see Villafranca papers.  
 \item $p$: Cluster membership probability.
 \item $\varpi$: Individual \textit{Gaia}~(E)DR3 parallax.  
 \item \pic: Individual corrected \textit{Gaia}~(E)DR3 parallax, see Eqn.~\ref{pic}.  
 \item \pig: Group parallax, see Villafranca papers.
 \item PA: Position angle in a binary system (secondary with respect to primary) from N towards E. 
 \item PDMF: Present Day Mass Function.
 \item PMS: Pre-main sequence.
 \item $r$: Filtering separation from the center in the plane of the sky, see Villafranca papers.
 \item \rc: Core radius for the King profile.
 \item \rmu: Filtering separation from the center in proper motion, see Villafranca papers. 
 \item \RV: Ratio of monochromatic extinction to color excess, see \citet{Maiz24}.
 \item RUWE: Renormalized Unit Weight Error.
 \item $\rho(r)$: Stellar number density.
 \item \sigmai: Individual \textit{Gaia}~(E)DR3 parallax/proper motion uncertainty, see Eqn.~\ref{sigmae}. 
 \item \sigmae: Individual \textit{Gaia}~(E)DR3 corrected parallax/proper motion uncertainty, see Eqn.~\ref{sigmae}. 
 \item \sigmas: \textit{Gaia}~(E)DR3 parallax/proper motion systematic uncertainty, see \citet{Maizetal21c} and Eqn.~\ref{sigmae}. 
 \item \spig: Group parallax uncertainty, see Villafranca papers.
 \item \sX: Astrophysical photometric dispersion for band $X$ (\GGG, \GGGBP, or \GGGRP) from \citet{Maizetal23}.
 \item SB1: Single-lined spectroscopic binary.
 \item SB2: Double-lined spectroscopic binary.
 \item SED: Spectral Energy Distribution.
 \item SLS: SMC Luminous Star.
 \item SMC: Small Magellanic Cloud.
 \item SN: Supernova.
 \item $t_\varpi$: Normalized $\chi^2$ test for the group parallax, see Villafranca papers.
 \item $t_{\mu_{\alpha *}}$: Normalized $\chi^2$ test for the group proper motion in $\alpha$, see Villafranca papers.
 \item $t_{\mu_{\delta}}$: Normalized $\chi^2$ test for the group proper motion in $\delta$, see Villafranca papers.
 \item $U_\odot$, $V_\odot$, $W_\odot$: Components of the peculiar solar velocity with respect to the LSR.
 \item \chired: Reduced $\chi^2$ of the CHORIZOS fit.
 \item WDS: Washington Double Star catalog, \citet{Masoetal01}.
 \item $z_\odot$: Height of the Sun above the Galactic Plane.
 \item ZP: Parallax zero point, see Eqn.~\ref{pic}.
\end{itemize}

\end{appendix}

\end{document}